\documentclass[11pt]{article}

\newif\ifcomments

\commentstrue

\usepackage{fullpage}
\usepackage[utf8]{inputenc}
\usepackage{tikz}
\usetikzlibrary{shapes.geometric, arrows}
\usepackage{graphicx}
\usepackage{amsmath,amsthm,amssymb}
\usepackage{algorithm}
\usepackage{algpseudocode}
\usepackage{multirow}
\usepackage{makecell}
\usepackage{algpseudocode}

\usepackage{thm-restate,color,xspace}
\usepackage{comment}
\usepackage{thmtools}
\usepackage{xcolor}
\usepackage{nameref}
\usepackage{array}
\usepackage{svg}

\definecolor{ForestGreen}{rgb}{0.1333,0.5451,0.1333}
\definecolor{DarkRed}{rgb}{0.65,0,0}
\definecolor{Red}{rgb}{1,0,0}
\usepackage[linktocpage=true,
pagebackref=true,colorlinks,
linkcolor=DarkRed,citecolor=ForestGreen,
bookmarks,bookmarksopen,bookmarksnumbered]
{hyperref}
\usepackage{cleveref}

\usepackage{xcolor}
\usepackage{nameref}

\usepackage{cleveref}

\makeatother

\usepackage{babel}
\usepackage{hyperref}

\declaretheorem[numberwithin=section]{theorem}
\declaretheorem[numberlike=theorem]{lemma}

\declaretheorem[numberlike=theorem]{property}

\declaretheorem[numberlike=theorem,style=definition]{remark}

\global\long\def\poly{\mathrm{poly}}

\newcommand{\ignore}[1]{}

\usepackage{todonotes}

\ifcomments
\def\thatchaphol#1{\marginpar{$\leftarrow$\fbox{TS}}\footnote{$\Rightarrow$~{\sf\textcolor{purple}{#1 --Thatchaphol}}}}
\def\benyu#1{\marginpar{$\leftarrow$\fbox{BW}}\footnote{$\Rightarrow$~{\sf\textcolor{red}{#1 --Benyu}}}}
\def\aditya#1{\marginpar{$\leftarrow$\fbox{GH}}\footnote{$\Rightarrow$~{\sf\textcolor{blue}{#1 --Aditya}}}}

\else
\newcommand{\benyu}[1]{}
\newcommand{\thatchaphol}[1]{}
\newcommand{\aditya}[1]{}
\fi 

\usepackage[utf8]{inputenc} 
\DeclareUnicodeCharacter{1D405}{$\mathbf{F}$} 

\DeclareUnicodeCharacter{1D400}{\ensuremath{\mathbf{A}}}
\DeclareUnicodeCharacter{1D401}{\ensuremath{\mathbf{B}}}
\DeclareUnicodeCharacter{1D402}{\ensuremath{\mathbf{C}}}
\DeclareUnicodeCharacter{1D403}{\ensuremath{\mathbf{D}}}
\DeclareUnicodeCharacter{1D404}{\ensuremath{\mathbf{E}}}
\DeclareUnicodeCharacter{1D405}{\ensuremath{\mathbf{F}}}
\DeclareUnicodeCharacter{1D406}{\ensuremath{\mathbf{G}}}
\DeclareUnicodeCharacter{1D407}{\ensuremath{\mathbf{H}}}
\DeclareUnicodeCharacter{1D408}{\ensuremath{\mathbf{I}}}
\DeclareUnicodeCharacter{1D409}{\ensuremath{\mathbf{J}}}
\DeclareUnicodeCharacter{1D40A}{\ensuremath{\mathbf{K}}}
\DeclareUnicodeCharacter{1D40B}{\ensuremath{\mathbf{L}}}
\DeclareUnicodeCharacter{1D40C}{\ensuremath{\mathbf{M}}}
\DeclareUnicodeCharacter{1D40D}{\ensuremath{\mathbf{N}}}
\DeclareUnicodeCharacter{1D40E}{\ensuremath{\mathbf{O}}}
\DeclareUnicodeCharacter{1D40F}{\ensuremath{\mathbf{P}}}
\DeclareUnicodeCharacter{1D410}{\ensuremath{\mathbf{Q}}}
\DeclareUnicodeCharacter{1D411}{\ensuremath{\mathbf{R}}}
\DeclareUnicodeCharacter{1D412}{\ensuremath{\mathbf{S}}}
\DeclareUnicodeCharacter{1D413}{\ensuremath{\mathbf{T}}}
\DeclareUnicodeCharacter{1D414}{\ensuremath{\mathbf{U}}}
\DeclareUnicodeCharacter{1D415}{\ensuremath{\mathbf{V}}}
\DeclareUnicodeCharacter{1D416}{\ensuremath{\mathbf{W}}}
\DeclareUnicodeCharacter{1D417}{\ensuremath{\mathbf{X}}}
\DeclareUnicodeCharacter{1D418}{\ensuremath{\mathbf{Y}}}
\DeclareUnicodeCharacter{1D419}{\ensuremath{\mathbf{Z}}}

\DeclareUnicodeCharacter{1D41A}{\ensuremath{\mathbf{a}}}
\DeclareUnicodeCharacter{1D41B}{\ensuremath{\mathbf{b}}}
\DeclareUnicodeCharacter{1D41C}{\ensuremath{\mathbf{c}}}
\DeclareUnicodeCharacter{1D41D}{\ensuremath{\mathbf{d}}}
\DeclareUnicodeCharacter{1D41E}{\ensuremath{\mathbf{e}}}
\DeclareUnicodeCharacter{1D41F}{\ensuremath{\mathbf{f}}}
\DeclareUnicodeCharacter{1D420}{\ensuremath{\mathbf{g}}}
\DeclareUnicodeCharacter{1D421}{\ensuremath{\mathbf{h}}}
\DeclareUnicodeCharacter{1D422}{\ensuremath{\mathbf{i}}}
\DeclareUnicodeCharacter{1D423}{\ensuremath{\mathbf{j}}}
\DeclareUnicodeCharacter{1D424}{\ensuremath{\mathbf{k}}}
\DeclareUnicodeCharacter{1D425}{\ensuremath{\mathbf{l}}}
\DeclareUnicodeCharacter{1D426}{\ensuremath{\mathbf{m}}}
\DeclareUnicodeCharacter{1D427}{\ensuremath{\mathbf{n}}}
\DeclareUnicodeCharacter{1D428}{\ensuremath{\mathbf{o}}}
\DeclareUnicodeCharacter{1D429}{\ensuremath{\mathbf{p}}}
\DeclareUnicodeCharacter{1D42A}{\ensuremath{\mathbf{q}}}
\DeclareUnicodeCharacter{1D42B}{\ensuremath{\mathbf{r}}}
\DeclareUnicodeCharacter{1D42C}{\ensuremath{\mathbf{s}}}
\DeclareUnicodeCharacter{1D42D}{\ensuremath{\mathbf{t}}}
\DeclareUnicodeCharacter{1D42E}{\ensuremath{\mathbf{u}}}
\DeclareUnicodeCharacter{1D42F}{\ensuremath{\mathbf{v}}}
\DeclareUnicodeCharacter{1D430}{\ensuremath{\mathbf{w}}}
\DeclareUnicodeCharacter{1D431}{\ensuremath{\mathbf{x}}}
\DeclareUnicodeCharacter{1D432}{\ensuremath{\mathbf{y}}}
\DeclareUnicodeCharacter{1D433}{\ensuremath{\mathbf{z}}}

\DeclareUnicodeCharacter{1D434}{\ensuremath{\mathit{A}}}
\DeclareUnicodeCharacter{1D435}{\ensuremath{\mathit{B}}}
\DeclareUnicodeCharacter{1D436}{\ensuremath{\mathit{C}}}
\DeclareUnicodeCharacter{1D437}{\ensuremath{\mathit{D}}}
\DeclareUnicodeCharacter{1D438}{\ensuremath{\mathit{E}}}
\DeclareUnicodeCharacter{1D439}{\ensuremath{\mathit{F}}}
\DeclareUnicodeCharacter{1D43A}{\ensuremath{\mathit{G}}}
\DeclareUnicodeCharacter{1D43B}{\ensuremath{\mathit{H}}}
\DeclareUnicodeCharacter{1D43C}{\ensuremath{\mathit{I}}}
\DeclareUnicodeCharacter{1D43D}{\ensuremath{\mathit{J}}}
\DeclareUnicodeCharacter{1D43E}{\ensuremath{\mathit{K}}}
\DeclareUnicodeCharacter{1D43F}{\ensuremath{\mathit{L}}}
\DeclareUnicodeCharacter{1D440}{\ensuremath{\mathit{M}}}
\DeclareUnicodeCharacter{1D441}{\ensuremath{\mathit{N}}}
\DeclareUnicodeCharacter{1D442}{\ensuremath{\mathit{O}}}
\DeclareUnicodeCharacter{1D443}{\ensuremath{\mathit{P}}}
\DeclareUnicodeCharacter{1D444}{\ensuremath{\mathit{Q}}}
\DeclareUnicodeCharacter{1D445}{\ensuremath{\mathit{R}}}
\DeclareUnicodeCharacter{1D446}{\ensuremath{\mathit{S}}}
\DeclareUnicodeCharacter{1D447}{\ensuremath{\mathit{T}}}
\DeclareUnicodeCharacter{1D448}{\ensuremath{\mathit{U}}}
\DeclareUnicodeCharacter{1D449}{\ensuremath{\mathit{V}}}
\DeclareUnicodeCharacter{1D44A}{\ensuremath{\mathit{W}}}
\DeclareUnicodeCharacter{1D44B}{\ensuremath{\mathit{X}}}
\DeclareUnicodeCharacter{1D44C}{\ensuremath{\mathit{Y}}}
\DeclareUnicodeCharacter{1D44D}{\ensuremath{\mathit{Z}}}

\DeclareUnicodeCharacter{1D44E}{\ensuremath{\mathit{a}}}
\DeclareUnicodeCharacter{1D44F}{\ensuremath{\mathit{b}}}
\DeclareUnicodeCharacter{1D450}{\ensuremath{\mathit{c}}}
\DeclareUnicodeCharacter{1D451}{\ensuremath{\mathit{d}}}
\DeclareUnicodeCharacter{1D452}{\ensuremath{\mathit{e}}}
\DeclareUnicodeCharacter{1D453}{\ensuremath{\mathit{f}}}
\DeclareUnicodeCharacter{1D454}{\ensuremath{\mathit{g}}}
\DeclareUnicodeCharacter{210E}{\ensuremath{\mathit{h}}}
\DeclareUnicodeCharacter{1D456}{\ensuremath{\mathit{i}}}
\DeclareUnicodeCharacter{1D457}{\ensuremath{\mathit{j}}}
\DeclareUnicodeCharacter{1D458}{\ensuremath{\mathit{k}}}
\DeclareUnicodeCharacter{1D459}{\ensuremath{\mathit{l}}}
\DeclareUnicodeCharacter{1D45A}{\ensuremath{\mathit{m}}}
\DeclareUnicodeCharacter{1D45B}{\ensuremath{\mathit{n}}}
\DeclareUnicodeCharacter{1D45C}{\ensuremath{\mathit{o}}}
\DeclareUnicodeCharacter{1D45D}{\ensuremath{\mathit{p}}}
\DeclareUnicodeCharacter{1D45E}{\ensuremath{\mathit{q}}}
\DeclareUnicodeCharacter{1D45F}{\ensuremath{\mathit{r}}}
\DeclareUnicodeCharacter{1D460}{\ensuremath{\mathit{s}}}
\DeclareUnicodeCharacter{1D461}{\ensuremath{\mathit{t}}}
\DeclareUnicodeCharacter{1D462}{\ensuremath{\mathit{u}}}
\DeclareUnicodeCharacter{1D463}{\ensuremath{\mathit{v}}}
\DeclareUnicodeCharacter{1D464}{\ensuremath{\mathit{w}}}
\DeclareUnicodeCharacter{1D465}{\ensuremath{\mathit{x}}}
\DeclareUnicodeCharacter{1D466}{\ensuremath{\mathit{y}}}
\DeclareUnicodeCharacter{1D467}{\ensuremath{\mathit{z}}}

\DeclareUnicodeCharacter{1D7CE}{\ensuremath{\mathbf{0}}}
\DeclareUnicodeCharacter{1D7CF}{\ensuremath{\mathbf{1}}}
\DeclareUnicodeCharacter{1D7D0}{\ensuremath{\mathbf{2}}}
\DeclareUnicodeCharacter{1D7D1}{\ensuremath{\mathbf{3}}}
\DeclareUnicodeCharacter{1D7D2}{\ensuremath{\mathbf{4}}}
\DeclareUnicodeCharacter{1D7D3}{\ensuremath{\mathbf{5}}}
\DeclareUnicodeCharacter{1D7D4}{\ensuremath{\mathbf{6}}}
\DeclareUnicodeCharacter{1D7D5}{\ensuremath{\mathbf{7}}}
\DeclareUnicodeCharacter{1D7D6}{\ensuremath{\mathbf{8}}}
\DeclareUnicodeCharacter{1D7D7}{\ensuremath{\mathbf{9}}}
\DeclareUnicodeCharacter{2265}{\ensuremath{\geq}}     
\DeclareUnicodeCharacter{2212}{\ensuremath{-}}        
\DeclareUnicodeCharacter{2032}{\ensuremath{^\prime}}  
\DeclareUnicodeCharacter{2113}{\ensuremath{\ell}}     
\DeclareUnicodeCharacter{22EF}{\ensuremath{\cdots}}   
\DeclareUnicodeCharacter{2033}{\ensuremath{^{\prime\prime}}}   
\DeclareUnicodeCharacter{2034}{\ensuremath{^{\prime\prime\prime}}} 

\algnewcommand{\algorithmicgoto}{\textbf{go to}}%
\algnewcommand{\Goto}[1]{\algorithmicgoto~\ref{#1}}%
\begin{document}

\title{
Strong Connectivity Oracle for Multiple Failures\\ via Rooted Flow-Augmentation}

\author{
  Aditya Anand\thanks{\texttt{adanand@umich.edu}.}
  \and Thatchaphol Saranurak\thanks{\texttt{thsa@umich.edu}. Supported by NSF Grant CCF-2238138.}
  \and Benyu Wang\thanks{\texttt{benyuw@umich.edu}.}
}

\date{}
\maketitle
\pagenumbering{gobble}

\begin{abstract}

We present a significantly improved  \emph{$k$-fault-tolerant strong connectivity oracle}, one of the most basic data structures on a directed graph $G$  that, given any set $F$ of failed edges of size $|F|\le k$, answers whether $G \setminus F$ is strongly connected. 
For every constant $k$, our oracle only uses $n^{1+o(1)}$ space and $n^{o(1)}$ query time. No prior result can achieve $o(n^{2})$ space and $o(n)$ query time, \emph{even when $k=2$}. 
Previous oracles either require $\Omega(n)$ query time \cite{baswana2018fault}, or $\Omega(n^2)$ space \cite{van2019sensitive}.


Our main technical contribution is a \emph{rooted} version of the highly successful flow augmentation technique, which may be of independent interest. Given a directed graph $G$, an integer $k$, and a source vertex $s$, we can sample a set of edges $A$ such that each minimal $s$-rooted cut of size at most $k$ becomes an $s$-rooted min-cut in $G + A$ with probability $2^{-\poly(k)}n^{-o(1)}$. 



\end{abstract}

\newpage
\tableofcontents
\newpage

\pagenumbering{arabic}
\section{Introduction}

A \emph{$k$-fault-tolerant strong connectivity oracle} is one of the most basic data structures in directed graphs: given a directed graph $G$, preprocess it into a data structure that, for any failure set $F$ of at most $k$ edges, answers whether $G \setminus F$ is still strongly connected. The goal is to minimize both the space and the query time.

When one optimizes only space or only query time, this problem is well-understood.
Optimizing only space, there is an oracle with $O(n2^k)$ space and linear query time~\cite{baswana2018fault}, obtained by storing the union of single-source reachability preservers of $G$ and of the reverse graph of $G$. This space complexity is essentially optimal: an almost tight lower bound of $\Omega(\min\{n^2,n2^{k/2}\})$ is implicit in~\cite{hoppenworth2025near}.
Optimizing only query time, there is an oracle with $O(n^2 \log n)$ space and $O(k^3)$ query time, obtained by building the all-pairs reachability oracle of~\cite{van2019sensitive} and, after deleting $k$ failed edges, checking reachability only between the endpoints of those failed edges.
Thus, for constant $k$, we can achieve \emph{either} linear space or constant query time.

Unfortunately, virtually no prior solution optimizes both simultaneously.\footnote{For $k=1$, there is a trivial oracle with $O(n)$ space and $O(1)$ query time: for each edge $e$ in the $O(n)$-size preserver of~\cite{baswana2018fault}, one simply marks whether $G \setminus \{e\}$ remains strongly connected.}
Even for $k=2$, no previous result gives both $o(n^2)$ space and $o(n)$ query time. This leaves the following basic question:
\begin{center}
\emph{For constant $k$, is there a strong connectivity oracle with $o(n^2)$ space and $o(n)$ query time?}
\end{center}

In this paper, we answer this question in a strong sense: for every constant $k$, we obtain an oracle with almost-linear space and almost-constant query time.

\begin{restatable}{theorem}{sccoracle}
\label{thm:sccoracle}\label{thm:intro-main}
    Given a directed graph $G$ and a positive integer $k$, there is an oracle that uses $2^{\poly(k)}n^{1 + o(1)}$ space, and given any set $F$ of at most $k$ edges, answers if $G \setminus F$ is strongly connected with high probability in time $2^{\poly(k)}n^{o(1)}$. 
\end{restatable}

Furthermore, given another positive parameter $\lambda$, we can strengthen the oracle to answer whether $G \setminus F$ is $\lambda$-strongly connected; see \Cref{thm:connoracle}. We note that even for this variant, it is unknown whether there is an $O(n^2)$-space oracle with $o(n)$ query time for constant $\lambda$ and $k$, since existing fast directed reachability oracles answer only pairwise reachability queries.

\paragraph{Strong connectivity is harder than $(s,t)$-reachability.}
A central difficulty is that strong connectivity is a \emph{global} property.
A simple reduction shows that the fault-tolerant strong connectivity oracle problem is at least as hard as the fault-tolerant \emph{$(s,t)$-reachability} oracle problem, which asks whether $s$ can still reach $t$ after failures.\footnote{Given $G$ and vertices $s,t$, add edges $(v,s)$ and $(t,v)$ for every $v \in V(G)\setminus\{s,t\}$ to obtain a graph $H$. Then $H \setminus F$ is strongly connected if and only if $t$ is reachable from $s$ in $G \setminus F$.}
The key difference is that a strong connectivity oracle must detect whether \emph{some} pair becomes unreachable, while an $(s,t)$-reachability oracle may focus on one prescribed pair.
This is why the recent $(s,t)$-reachability oracle of~\cite{ahi2026maximum}, with $O(2^{\poly(k)} n \log n)$ space and $O(2^{\poly(k)} \log n)$ query time, does not extend directly to our setting.

\paragraph{Our main technical contribution: rooted flow augmentation.}
To obtain Theorem~\ref{thm:intro-main}, we develop a new \emph{rooted} version of flow augmentation. Flow augmentation has recently emerged as a powerful primitive in parameterized graph algorithms~\cite{kim2021solving,kim2022directed,kim2024weighted,kim2025flow}. In its usual form, one is given a source $s$ and a \emph{fixed} sink $t$, and the goal is to add edges so that a target minimal $(s,t)$-cut becomes an $(s,t)$ min-cut.

For strong connectivity, this is not enough. The obstacle is exactly that the relevant sink side is \emph{not fixed in advance}: after failures, the witness to non-strong-connectivity could be any rooted minimal cut. Our first contribution is to remove this dependence on a prescribed sink and make flow augmentation work for \emph{rooted} cuts.

Recall that an $s$-rooted cut is a set of edges whose deletion prevents $s$ from reaching some vertex, and it is minimal if no proper subset has the same property. An $s$-rooted min-cut is simply an $s$-rooted cut of minimum capacity. Our rooted flow-augmentation theorem says that one can add edges so that any fixed rooted minimal cut of size at most $k$ becomes an $s$-rooted min-cut with probability $2^{-\poly(k)} n^{-o(1)}$.

\begin{restatable}{theorem}{strongaugment}
\label{thm:strong-augment}
\label{thm:intro-rooted}
There exists a randomized algorithm running in time $2^{\poly(k)}n^{O(1)}$ that, given a directed graph $G$, a root vertex $s \in V(G)$, and an integer $k$, outputs a set $A \subseteq V(G) \times V(G)$ such that the following holds: for every $s$-rooted minimal cut $Z \subseteq E(G)$ of size at most $k$, with probability $2^{-\poly(k)} n^{-o(1)}$, $Z$ becomes an $s$-rooted min-cut in $G + A$.
\end{restatable}

Following \cite{kim2022directed}, by $G + A$, we mean the graph $G$ with adding all edges in $A$ with infinite capacity.\footnote{To make the graph unweighted, we can replace each infinite capacity edge in $A$ with $k+1$ parallel edges.} The $\poly(k)$ factor in the exponent is the same as that in the $s$-$t$ flow-augmentation result of~\cite{kim2022directed}. Currently, it is $O(k^4 \log k)$, and any improvement to~\cite{kim2022directed} would immediately improve our result as well.

The $s$-$t$ flow-augmentation result of~\cite{kim2022directed} has turned out to be a powerful primitive in parameterized graph algorithms~\cite{kim2021solving,kim2022directed,kim2024weighted,kim2025flow}. Therefore, we believe Theorem~\ref{thm:intro-rooted}, as a generalization from $s$-$t$ cuts to global cuts, could be of independent interest.

\paragraph{The many-graph interface.}
For the oracle, the most convenient form of rooted flow augmentation is a \emph{many-graph} version.
Roughly speaking, instead of forcing a rooted minimal cut to become a rooted min-cut inside the original graph, we localize it into one graph in a family whose \emph{total} size remains near-linear. In that graph, the cut becomes a standard $(s,t_i)$ min-cut for a designated sink $t_i$.
This is exactly the form needed to plug in the recent fault-tolerant $(s,t)$-reachability oracle of~\cite{ahi2026maximum} as a black box.

\begin{restatable}[Many-graph version of rooted flow augmentation]{theorem}{weakaugment}
\label{thm:weak-augment}
\label{thm:intro-many-graph}
There exists a randomized algorithm running in time $2^{\poly(k)}n^{O(1)}$ that, given a directed graph $G=(V,E)$, a root vertex $s \in V$, and an integer $k$, outputs a subpartition\footnote{A subpartition of $V$ is a collection of pairwise disjoint subsets of $V$.} $\mathcal{R} = \{R_i\}_{i \in [p]}$ of $V$, a sink set $\mathcal{T} = \{t_i\}_{i \in [p]}$, and edge sets $\mathcal{A} = \{A_i\}_{i \in [p]}$, where $A_i \subseteq (R_i \cup \{s\}) \times R_i$. Let $G_i$ be the graph obtained by contracting $V \setminus R_i$ to the source $s$.

Then any $s$-rooted minimal cut $Z \subseteq E$ of size at most $k$ becomes an $s$-rooted min-cut in $G_i + A_i$, with $t_i$ on the sink side, for some $i \in [p]$, with probability at least $2^{-\poly(k)} n^{-o(1)}$.
\end{restatable}

We remark that $\sum_i |E(G_i)| \le |E(G)|$ and $\sum_i |V(G_i)| = O(|V(G)|)$ since $\mathcal{R}$ is a subpartition.

Theorem~\ref{thm:intro-many-graph} is the theorem that interfaces directly with the oracle. Once we have it, we can build a collection of pointwise $(s,t_i)$ failure oracles on the graphs $G_i$, repeat the randomized construction a small number of times, and obtain Theorem~\ref{thm:intro-main}. Theorem~\ref{thm:intro-rooted} is stronger as a standalone graph-theoretic statement, but Theorem~\ref{thm:intro-many-graph} is more convenient for obtaining our oracle.\footnote{One can also derive Theorem~\ref{thm:intro-main} from Theorem~\ref{thm:intro-rooted} using \cite{gabow1995centroids} and \cite{ahi2026maximum}, but in a less direct way.}

\paragraph{Roadmap.}
\Cref{sec:overview} gives the technical overview. \Cref{sec:prelim} collects preliminaries. In \Cref{sec:augment}, we first prove the many-graph rooted flow-augmentation theorem, which is the key input to the oracle, and then strengthen it to rooted flow augmentation in the original graph. \Cref{sec:oracle} combines the many-graph theorem with recent $(s,t)$ sensitivity oracles to obtain our strong connectivity oracle, and also extends the result to $\lambda$-strong connectivity.

\paragraph{Undirected graphs.}
For comparison, the analogous problem in undirected graphs is much better understood: there are $k$-fault-tolerant connectivity oracles with $\tilde{O}(n)$ space and $\tilde{O}(k)$ query time, and these results extend to the harder fully dynamic setting~\cite{kapron2013dynamic}. Our result shows that for directed \emph{strong} connectivity, one can now obtain a similarly efficient fault-tolerant oracle for every constant number of failures.

\paragraph{More related work.}
\emph{fault-tolerant strongly connected components (SCCs) oracle} is very closely related to our problem on fault-tolerant strong connectivity. The goal is to report all SCCs instead of solving the decision version whether the graph remains strongly connected.
This problem has been studied extensively.
For a single failure, Georgiadis, Italiano, and Parotsidis~\cite{GIP17} give a linear-space oracle that reports all SCCs in linear time.
For multiple failures, Baswana, Choudhary, and Roditty~\cite{BCR17SCC} give a fault-tolerant SCC oracle, and subsequent work developed increasingly sparse fault-tolerant strong-connectivity preservers~\cite{CC20,hoppenworth2025near}.
These results, however, do not improve the previously known space--query tradeoff from \cite{baswana2018fault} for the decision problem considered here.

\emph{Dynamic algorithms for strong connectivity}, which handle an online sequence of updates, has also been studied extensively.
In the \emph{incremental} setting, a long line of work on maintaining SCCs and detecting directed cycles~\cite{HKMST12,BFGT16,BC18,BDP21} culminated in a deterministic $m^{1+o(1)}$ total-update-time algorithm based on interior-point methods~\cite{CKLMP24}.
In the \emph{decremental} setting, a long line of work~\cite{HKN14,HKN15,CHILP16,BPW19,BGS20} culminated in a deterministic $m^{1+o(1)}$ total-update-time algorithm based on interior-point methods~\cite{BCKLMPS24}.
In the \emph{fully dynamic} setting, strong connectivity has been studied both directly and through dynamic reachability and matrix inverse~\cite{San04,BNS19,KS23}, but the update time is at least $O(n^{1.406})$.

Together, these works further motivate our fault-tolerant strong connectivity as a basic and extensively studied problem, while leaving open the space--query tradeoff addressed in this paper.

\section{Technical Overview}
\label{sec:overview}

\paragraph{Strong Connectivity Oracle}
First, we show that one can essentially reduce constructing a strong connectivity oracle to our flow augmentation result (\Cref{thm:weak-augment}). The statement of \Cref{thm:weak-augment} implies that one can find graphs $G_i$, and sinks $t_i \in V(G_i)$, and a random set of new edges ${A}_i$ with the guarantee that any fixed $s$-rooted minimal cut $Z$ becomes a $s$-$t_i$ min-cut in $G_i$ with probability $2^{-\poly(k)}n^{-o(1)}$ after addition of the edges ${A}_i$, for some $i \in [p]$. If we independently repeat the algorithm of \Cref{thm:weak-augment} $2^{\poly(k)}n^{o(1)}$ times, then for any cut $Z$, with high probability, in at least one of the random repetitions we can ensure that $Z$ is an $s$-$t_i$ min-cut for some $i \in [p]$. 


Since we know the set of $t_i$, the graphs $G_i$, and the sets ${A}_i$ we can simply construct the $s$-$t$ oracle of~\cite{ahi2026maximum} for $s$-$t_i$ min-cuts of size at most $k$ for each $t_i$ in the graph $G_i + A_i$. The result of~\cite{ahi2026maximum} would take $G_i + A_i,s_i,t_i$ and construct an oracle that can answer $s$-$t_i$ min-cut queries after the failure of any at most $k$ edges (In particular, we can answer $s$-$t_i$ reachability queries, which just amounts to checking if the min-cut is zero). The crucial property about \Cref{thm:weak-augment} is that the total size of all graphs $G_i$ is small: they have at most $O(m)$ edges and $O(n)$ vertices, which now helps us to bound the total space and query time (note that we repeat \Cref{thm:weak-augment} only $2^{\poly(k)}n^{o(1)}$ times to boost the probability). Finally, as mentioned earlier in the introduction, recall that one can assume that $m = O(n \cdot 2^k)$ by sparsifying the input graph using the result of~\cite{baswana2018fault}.  

Thus, it is sufficient to prove \Cref{thm:weak-augment}. The rest of the overview will focus on the proof of \Cref{thm:weak-augment}, which is our key technical contribution. At the end,  we briefly mention our ideas to obtain the stronger result of \Cref{thm:strong-augment}.

We note here that there is an alternative approach to obtaining the oracle from the structural decomposition that we obtain to get \Cref{thm:weak-augment}. It can be shown that in our construction, the set of $t_i$ in \Cref{thm:weak-augment} hit the non $s$-side of every rooted minimal cut, thus we could directly call the result of~\cite{ahi2026maximum} for $s$-$t_i$ failures of size at most $k$ in $G_i$, removing the need for augmenting flow via $A_i$. However while this seems simpler, there is essentially no difference in these two approaches, as the first step of~\cite{ahi2026maximum} is to perform $s$-$t_i$ flow augmentation anyway. We choose to present rooted flow augmentation (\Cref{thm:weak-augment}) as an independent subroutine because we believe it could be a useful primitive in its own right and our algorithm/decomposition do not become any more complex due to this fact.   



 \paragraph{Rooted Flow-Augmentation} We will need the key tool of important separators, so we define it before moving on to our algorithm. In this paper, by an $(s,t)$ separator (or an $(s,t)$ cut), we mean a set of edges $X$ after whose deletion, there is no path from $s$ to $t$. We say that $X$ is minimal if no $X' \subsetneq X$ is an $(s,t)$ separator. An $(s,t)$ important separator\footnote{Note that our definition is slightly different from the usual notion of important separators for the ease of presentation.} is a minimal $(s,t)$ separator $X$, so that if $R(X)$ is the set of vertices that have a directed path to $t$ in $G \setminus X$, then for every $(s,t)$ separator $X'$ with $|X'| \leq |X|$, $R(X') \not\supset R(X)$. Thus these are the cuts closest to $s$ in terms of (inclusion-wise) maximizing the set of vertices that can reach $t$. It is a standard result in parameterized complexity that there are at most $4^k$ $(s,t)$ important separators of size at most $k$~\cite{chen2009improved,marx2006parameterized}.

 We also need the $(s,t)$ flow augmentation algorithm from~\cite{kim2022directed}: given a graph $G$, fixed vertices $s$ and $t$, and an integer $k$, we can add in polynomial time a random set of edges $A$ such that for any minimal $(s,t)$ cut $Z$ of size at most $k$, $Z$ becomes an $s$-$t$ minimum cut in $G + A$ with probability $2^{-\poly(k)}$. Whenever we write $G + A$, we mean $G$ with the edges in $A$ added with \emph{infinite} capacity.

Let us start by describing our algorithm for the many-graph version of rooted flow augmentation, \Cref{thm:weak-augment}. We start with a very simple but important observation about the structure of rooted minimal cuts. Recall that a subset $Z \subseteq E(G)$ is an $s$-rooted minimal cut if no proper subset of $Z$ is an $s$-rooted cut. For a set $T$, let $\delta^{-}(T)$ denote the set of incoming edges to $T$. We claim that $Z = \delta^{-}(T)$ for some \emph{strongly connected set} of vertices $T$. This can be seen as follows. Let $U$ be the set of vertices not reachable from $s$ after deleting $Z$. If $U$ is strongly connected, then we are done, and we let $T = U$. Otherwise, we let $T$ be the first SCC in any topological ordering of $G[U]$. It is easy to see that minimality forces $Z = \delta^{-}(T)$, for otherwise, we can remove the edges $Z \setminus \delta^{-}(T)$ to still form an $s$-rooted cut. Thus equivalently, one can represent any $s$-rooted minimal cut $Z$ by a set of vertices $T$ that is strongly connected.

Fix some parameter $\theta$ that will be determined by the analysis. First, we try to capture all such sets $T$ that are \emph{large} with $|T| \geq \frac{n}{\theta}$. If we sample $ \approx k\theta \log n$ vertices $Q$, we can prove that $Q$ is a hitting set for every set of vertices $T$\footnote{Technically, this only holds if $T$ is an in-reachable cut, such that every vertex of $T$ can reach a common sink $t \in T$, but we skip this for simplicity in this overview.} such that $|\delta^-(T)| \leq k$ and $|T| > \frac{n}{\theta}$ with high probability. This can be seen via a simple Chernoff bound: the expected value of $|Q \cap T|$ is $\approx  k \log n$, which yields the right concentration to union bound over $\approx n^{2k}$ cuts of size $k$. For simplicity and ease of exposition in this overview, we assume that $Q$ is a hitting set for every such $T$ with probability $1$ (whereas it only holds with high probability).

Now sample $q \in Q$ uniformly at random, and do $(s,t)$ flow augmentation with $t = q$ to get a set of edges $A$. Then we can guarantee that for any minimal cut $Z$ with $Z = \delta^{-}(T)$ for some $|T| > \frac{n}{\theta}$, with probability $\frac{1}{|Q|}2^{-\poly(k)}$, $Z$ becomes an $(s,q)$ min-cut in $G + A$. If we then add edges $(q,v)$ for every vertex $v \in V(G)\setminus \{s\}$ to get a set of edges $B_q$, then we can additionally ensure that $Z$ is an $s$-rooted min-cut: any $s$-rooted min-cut cannot contain $q$ on the $s$-side due to the arcs $B_q$. Thus $Z$ becomes an $s$-rooted min-cut in $G + A + B_q$ with probability $\frac{1}{|Q|}2^{-\poly(k)}$ (Recall that in $G + A + B_q$, the edges $A \cup B_q$ have infinite capacity).

Now it remains to handle cuts $Z$ that correspond to sets $T$ which are of size less than $\frac{n}{\theta}$. We will construct a family of sets $\mathcal{C}$ as follows. Contract the set $Q$ into the vertex $s$. For simplicity, we still call the contracted vertex $s$. Next for every vertex $v \neq s$, we compute the set of all $(s,v)$ important separators of size at most $k$.

Consider any such $(s,v)$ important separator $X = \delta^{-}(T_v)$. Let $C_v \subseteq T_v$ be the vertex set of the SCC containing $v$ inside $T_v$. We then let $\mathcal{C}$ be the collection of all (inclusion-wise maximal) $C_v$ across all $v \in V(G) \setminus (Q \cup \{s\})$.

By the definition of $(s,v)$ important separators (they are the separators ``closest'' to $s$) and the fact that the non-$s$ side $T$ of a minimal cut is strongly connected, it is easy to show that every minimal cut $Z = \delta^{-}(T)$ must satisfy $T \subseteq C$ for some $C \in \mathcal{C}$. Since we already hit all the large sets with cut-size at most $k$ and the set $Q$ was contracted into $s$, we know that $|T_v| < \frac{n}{\theta}$ for every $v$, and hence $|C| < \frac{n}{\theta}$ for every $C \in \mathcal{C}$. Next, we show (in \Cref{lem:impscc-partition}) that the family of sets $\mathcal{C}$ can be partitioned into at most $k\cdot4^k$ sub-families $\mathcal{C}_i$, such that for any $i \in [k\cdot4^k]$, the sets $C \in \mathcal{C}_i$ are vertex disjoint. In this overview, we skip the proof of this lemma, but the high-level idea is that important separators do not overlap too much. Once we build such families $\mathcal{C}_i$, pick one $\mathcal{C}_i$ uniformly at random. Then for each $C \in \mathcal{C}_i$, one by one, we contract everything outside $C$ into $s$ to obtain a new graph, in which we recursively run rooted flow-augmentation. That concludes the rough outline of our algorithm.

Now let us to try to analyze the probability with which a fixed minimal cut $Z$ becomes a minimum cut at some level of this recursive procedure.
The key property is that any minimal cut $Z = \delta^{-}(T)$ satisfies either of the following.

\begin{itemize}
\item $T \cap Q \neq \emptyset$. In this case, it is captured with probability ${2^{-\poly(k)}}{|Q|}^{-1}$ since we sample a random $q \in Q$, and perform $(s,q)$ flow augmentation.

\item $|T| < \frac{n}{\theta}$. In this case, with probability at least $\frac{1}{k4^{k}}$, we choose the family $\mathcal{C}_i$ so that $T \subseteq C$ for some $C \in \mathcal{C}_i$. This cut will then be captured while recursively running flow augmentation when we contract $V(G) \setminus C$. 
\end{itemize}

Whenever the algorithm makes a recursive call, we reduce the number of vertices from $n$ to $\frac{n}{\theta}$, thus the recursion depth is at most $L \approx \log_\theta n$. For a given cut $Z = \delta^{-}(T)$, with probability at least $(k \cdot 4^k)^{-L} =  2^{-O(Lk)}$, we reach some graph $H$ where $Z$ is still an $s$-rooted minimal cut, and the sampled set $Q$ satisfies $Q \cap T \neq \emptyset$. Choose $q \in Q \cap T$. $q$ must be sampled with probability $\frac{1}{|Q|} \approx \frac{1}{k\theta\log n}$. If $q$ is sampled, then the $(s,q)$ flow augmentation in $H$ returns a set of edges $A$. Adding the edges $A$ and the return arcs $B_q$ from $q$ to every vertex of $V(H) \setminus \{s\}$, $Z$ becomes an $s$-rooted min-cut in $H + A + B_q$ with probability $2^{-\poly(k)}$. Thus the overall probability of success is at least $2^{-\poly(k)}2^{-O(kL)}\frac{1}{k\theta\log n}$.

Set $\theta \approx 2^{\sqrt{k\log n}}$, then $L \approx \sqrt{\log n/k}$ - the success probability becomes $2^{-\poly(k)-O(\sqrt{k\log n})}$. By a standard argument, the term $2^{-O(\sqrt{k\log n})}$ is at least $2^{-\poly(k)} n^{-o(1)}$, since if $k \leq \sqrt{\log n}$, it is at least $n^{-o(1)}$, and otherwise, it is at least $2^{-\poly(k)}$.

\paragraph{Augmenting in the Original Graph $G$.} Now that we have the many-graph version of our flow augmentation, below, we briefly describe the challenge in extending the ideas above to obtain flow-augmentation in the original graph $G$ as in \Cref{thm:strong-augment}. Intuitively, one could try and follow the same procedure - but the following difficulty precludes the same analysis. Consider one sub-family $\mathcal{C}_i$ that we choose as in \Cref{thm:weak-augment}. The previous algorithm contracts, one by one, for each $C \in \mathcal{C}_i$ everything outside $C$ to obtain a new graph in which we perform recursive flow augmentation. However, if we want to add arcs to the original graph, then we must find a way to deal with all $C \in \mathcal{C}_i$ at once. Consider a minimal cut $Z = \delta^{-}(T)$ such that $T \subseteq C$ for some $C \in \mathcal{C}_i$. If we simply add all the resulting arcs $A_{C}$ from the recursive augmentation across all $C \in \mathcal{C}_i$ at the same time, then it is possible that $Z$ may not become a min-cut with significant probability: this is because \emph{the $s$-rooted min-cuts in the resulting graph may intersect more than one $\mathcal{C}_i$}. Such cuts could preclude the cut $Z$ from being a min-cut.

However, suppose after we add all the resulting arcs $A_{C}$ for each $C$, there is a $s$-rooted min-cut that intersects both $C_1$ and $C_2$ for two different  $C_1,C_2$ in $\mathcal{C}_i$. We show that such a scenario can only happen when there are edges between $C_1$ and $C_2$ - on a high level, if there are no edges between them, the individual flow augmentations $A_{C_1}$ and $A_{C_2}$ for $C_1$ and $C_2$ would have made this cut sufficiently large. Therefore, if somehow there were no edges between $C_1$ and $C_2$ for any two $C_1, C_2 \in \mathcal{C}_i$, we would obtain the desired outcome. We achieve this by showing a much stronger structural decomposition: in each recursion, we further refine the families $\mathcal{C}_i$ to obtain a new family of sets $\mathcal{D}_i$: every relevant $s$-rooted minimal cut is still contained in some $D_i \in \mathcal{D}_i$ for some $i$, and there are at most $k^2\cdot 4^k \mathrm{polylog}(n)$ such $i$ just like with $\mathcal{C}_i$ - but we now have the additional property that in every $\mathcal{D}_i$ and any two distinct $D_1, D_2 \in \mathcal{D}_i$, they are not just vertex disjoint, but there are also no edges between $D_1$ and $D_2$.


\section{Preliminaries}
\label{sec:prelim}
\paragraph{Graph Notions.} We consider a directed graph $G=(V,E)$ with $n$ vertices and $m$ edges. We use $V=V(G)$ and $E=E(G)$ to denote the vertex and edge sets in $G$. When $G$ has no cycle, we call $G$ a directed acyclic graph (DAG). All graphs below allow parallel edges and edge capacities can be $1$ or $+\infty$. For any edge set $Z \subseteq E$ with only capacity-$1$ edges, we define $G-Z$ to be the graph obtained by deleting $Z$ from $G$. Based on \cite{kim2022directed}, for a graph $G$ and an edge set $A \subseteq V \times V$, we define $G+A$ as the graph adding $A$ to $G$, where all edges in $G$ have capacity $1$ and all added edges in $A$ have capacity $+\infty$.

For any vertex set $S$ in graph $G$, let $G[S]$ be the subgraph of $G$ induced by vertices in $S$. We define $\delta_G^+(S)$ and $\delta_G^-(S)$ as the edges going out of $S$, or into $S$, respectively. When $S$ is a singleton $\{s\}$, we also write them as $\delta_G^+(s)$ or $\delta_G^-(s)$. We define $\rho_G(S)$ as the total capacity of edges in $\delta_G^-(S)$. We use $N_G^+(S)$ and $N_G^-(S)$ to denote the out- and in-neighbors of $S$. When the graph $G$ is clear in context, we omit the subscript $G$.

We say $G$ is strongly connected if for every pair of vertices $(u,v)$ in $G$ there exists a path from $u$ to $v$. We say $G$ is $\lambda$-(strongly) connected if for every pair of vertices $(u,v)$ in $G$ there exists $\lambda$ edge disjoint paths from $u$ to $v$, or equivalently, $\rho_G(S) \geq \lambda$ for all $S \subset V$. A vertex set $C$ is called a strongly connected component (SCC) in $G$ if $G[C]$ is strongly connected but there's no larger $C' \supseteq C$ that $G[C']$ is also strongly connected. Any graph can be viewed as a DAG of strongly connected components.

\paragraph{Cuts.}
For a graph $G$ and vertices $s,t$, an edge set $Z$ in $G$ is an $(s,t)$ cut if there is no path from $s$ to $t$ in $G-Z$, moreover, it is an $(s,t)$ \emph{minimal} cut if no proper subset of $Z$ is also an $(s,t)$ cut. An $(s,t)$ cut $Z$ is an $(s,t)$ \emph{minimum} cut (or an $(s,t)$ \emph{min-cut}) if it has minimum capacity among all $(s,t)$ cuts. An $(s,t)$ minimum cut is also minimal but not vice versa. All the definitions generalize to sets $(S,T)$ where we contract $S$ as source and $T$ as sink.

For any $(s,t)$ \emph{minimal} cut $Z$, it can be represented by a bipartition $(S,T)$ of all vertices, where $s \in S, t \in T$ and $Z = \delta^-(T)$. Below, we will use both an edge set $Z$ and a bipartition $(S,T)$ to represent cuts. When $(S,T)$ is used, we mean the edge set $\delta^-(T)$. An $(s,t)$ cut $(S,T)$ is \emph{out-reachable} if $s$ can reach every vertex in $S$ in $G-\delta^-(T)$, and is \emph{in-reachable} if every vertex in $T$ can reach $t$ in $G-\delta^-(T)$.

Similarly, for a graph $G$ and a rooted vertex $s$, an edge set $Z$ in $G$ is an $s$-rooted cut if there is no path from $s$ to \emph{some} vertex $v$ in $G-Z$, moreover, it is an $s$-rooted minimal cut if also no proper subset of $Z$ is an $s$-rooted cut. An $s$-rooted cut $Z$ is an $s$-rooted minimum cut (or an $s$-rooted \emph{min-cut}) if it has minimum capacity among all $s$-rooted cuts. An $s$-rooted minimum cut is also minimal but not vice versa. We say a graph is $s$-rooted $k$-connected if $s$-rooted min-cuts have capacity at least $k$.

We note that, any $s$-rooted minimal cut $Z$ is an $(s,t)$ minimal cut, for every vertex $t$ not reachable by $s$ in $G-Z$. On the other side, an $(s,t)$ minimal cut $Z$ may not be $s$-rooted minimal, since a proper subset of $Z$ may still disconnect other vertices than $t$ from $s$. For any $s$-rooted minimal cut $Z$, it can be represented by a bipartition $(S,T)$ of all vertices, where $s \in S$ and $Z = \delta^-(T)$. 

\paragraph{Flow-Augmentation} Below, we state the $(s,t)$ flow-augmentation theorem, as in \cite{kim2022directed}\footnote{The flow augmentation result has derandomization in its journal version \cite{kim2025flow1}. The reason we state the randomized version is that our approach seems intrinsically random and we state this for coherence.}.

\begin{theorem}[\cite{kim2022directed}]\label{thm:staugment}
There exists a randomized polynomial-time algorithm that, given a directed graph $G$, two vertices $s,t \in V(G)$, and an integer $k$, outputs a set $A \subseteq V(G) \times V(G)$ such that the following holds: for every minimal $(s,t)$-cut $Z \subseteq E(G)$ of size at most $k$, with probability $2^{-O(k^4 \log k)}$, $Z$ becomes a minimum $(s,t)$-cut in $G+A$.
\end{theorem}

Next, we state the notions of furthest (in-Reachable) min-cuts and important separators\footnote{We note that, for simplicity, these two notions are both stated in a reversed direction from the commonly used notations. To see the equivalency, just reverse all edges in the graph.}.

\paragraph{Furthest (in-Reachable) Min-Cuts.} It is folklore that a \emph{furthest} min-cut exists from the submodularity of cut values. Here, we state and prove the following version for the existence of a furthest \emph{in-reachable} min-cut.

\begin{lemma}[Folklore]
    \label{lem:fmc}
    Given two vertices $s,t$ in $G$, among all sets $T$ that $\delta^-(T)$ is an in-reachable min-cut, there exists a unique (inclusion-wise) maximal set $T$. We denote $\delta^-(T)$ (or $(V-T,T)$) as an $(s,t)$-furthest in-Reachable min-cut for such $T$.
\end{lemma}

\begin{proof}
    Suppose towards contradiction that there are two different maximal sets $T,T'$ where $\delta^-(T)$ and $\delta^-(T')$ are both \emph{in-reachable} min-cuts. From submodularity of cut values we have $\rho(T) + \rho(T') \geq \rho(T \cap T') + \rho(T \cup T')$. On the other side, we know $\delta^-(T)$ and $\delta^-(T')$ are both $(s,t)$ min-cuts and $\delta^-(T \cap T')$ and $\delta^-(T \cup T')$ are still valid $(s,t)$ cuts, so $\rho(T) + \rho(T') \leq \rho(T \cap T') + \rho(T \cup T')$. 
    
    Hence, $\delta^-(T \cap T')$ and $\delta^-(T \cup T')$ are both $(s,t)$ min-cuts. One can also observe that $(V-(T\cup T'),(T \cup T'))$ is an $(s,t)$ in-reachable cut since $T$ and $T'$ are both in-reachable. Therefore, $(V-(T\cup T'),(T \cup T'))$ is an $(s,t)$ in-reachable min-cut and hence both $T$ and $T'$ cannot be minimal, so there's a contradition and the statement holds.
\end{proof}

\paragraph{Important Separators.}
Next, we introduce the notion of \emph{important separators} \cite{marx2011important,cygan2015parameterized} as a generalization of the above definition. For any $(s,t)$ minimal cut $(S,T)$, we call it an \emph{important separator} if it is in-reachable, and no other in-reachable cut $(S',T')$ can have $\rho(T') \leq \rho(T)$ and also $T' \supset T$. That is, there is no in-reachable cut closer to $s$ with less or equal capacity. We state the following theorem about important separators:

\begin{lemma}[\cite{cygan2015parameterized}, Theorem 8.36]\label{lem:impsep}

For any graph $G$ and two vertices $s,t \in G$, let $\mathcal{T}$ be all $(s,t)$ important separators with capacity at most $k$, then $|\mathcal{T}| \leq 4^k$ and $\mathcal{T}$ can be constructed in time $O(|\mathcal{T}| \cdot k^2m)$.
    
\end{lemma}

\paragraph{Probability Bounds.} When we state an event happens \emph{with high probability}, we mean it can happen with probability $1-O(n^{-c})$ for any constant $c$ fixed before running the randomized algorithm. By union bound, suppose polynomially many events $\{X_i\}$ each happens with high probability, then with high probability all of them will happen. 

We state and use the following version of Chernoff bound:

\begin{lemma}[Chernoff Bound]\label{lem:chernoff}
    Let $\{X_i\}$ be independent random variables taking values in $\{0,1\}$ and $X = \sum_i X_i$. Let $\mu = E[X]$. We have:
    \[\Pr[|X-\mu| \geq \delta\mu] \leq 2e^{-\delta^2\mu/3}, \forall \delta \in [0,1].\]
\end{lemma}

\section{Rooted Flow-Augmentation}
\label{sec:augment}

Below, we prove \Cref{thm:strong-augment} and \Cref{thm:weak-augment}. Without loss of generality, we assume that $s$ can reach every vertex $v$ in the original graph $G$, otherwise, the empty cut is the only $s$-rooted minimal cut. We first prove a simple structural lemma that helps us to represent rooted minimal cuts. We show that for any $s$-rooted minimal cut $Z$, we can not only assume $Z=\delta^-(T)$ for some $T$, but also assume that the set $T$ is strongly connected. We state our observation as in the following lemma:

\begin{lemma}\label{lem:rootedrep}
    For any $s$-rooted minimal cut $Z$ in $G$, there exists some vertex set $C \not\ni s$ where $G[C]$ is strongly connected such that $Z = \delta^-(C)$. Hence minimal $Z$ can equivalently be represented by the bi-partition $(V-C,C)$ where $G[C]$ is strongly connected and $Z = \delta^-(C)$.
\end{lemma}

\begin{proof}
    Since $Z$ is minimal, we must have $Z=\delta^-(T)$ for some vertex set $T$ and $s \notin T$. If $T$ is strongly connected, then we take $C=T$. Otherwise, we decompose $T$ into its strongly connected components and let $C$ be the first component in topological order.

    We claim that $\delta^-(C)=\delta^-(T)$. Since $C$ is the first component in $G[T]$, we know $\delta^-(C)$ contains no edge from $T-C$ to $C$. Therefore, $\delta^-(C) \subseteq \delta^-(T)$. Then since $Z=\delta^-(T)$ is a $s$-rooted \emph{minimal} cut and $\delta^-(C)$ is also an $s$-rooted cut which disconnects $C$, we must have $\delta^-(C) = \delta^-(T) = Z$.
\end{proof}

\subsection{Many-Graph Version of Flow-Augmentation}

 We first prove the many-graph version of flow-augmentation, \Cref{thm:weak-augment}.

\weakaugment*

We briefly outline the proof strategy: Let $\theta$ be a fixed parameter to be determined. Based on \Cref{lem:rootedrep}, every $s$-rooted minimal cut $Z$ can be represented by some $(S,T)$, where $T$ is strongly connected and $Z = \delta^-(T)$. Suppose the component $T$ is large, so that satisfies $|T|>n/\theta$, we then capture $T$ by sampling some $t \in C$ and then use an $(s,t)$ flow-augmentation as in \Cref{thm:staugment}. Otherwise, we find a family $\{C_i\}$ that covers all possible small $T$'s, and then recurse in each $C_i$ to capture $T$. In this way, we first state a decomposition with some properties, then proceed to the flow-augmentation algorithm.

\subsubsection{Partitions}

We start by sampling each vertex with probability $\Theta(k\theta\log n/n)$ to form a set $Q = \{q_i\}$. Here, based on a standard Chernoff bound in \Cref{lem:chernoff}, we show the size of $Q$ is not too large with high probability.

\begin{property}
    \label{ass:samplesize}
    $|Q|=O(k\theta\log n)$.
\end{property}

\begin{lemma}
    \Cref{ass:samplesize} holds with high probability.
\end{lemma}

\begin{proof}
   Since we sample each vertex with probability $\Theta(k\theta\log n/n)$, the expected size of $Q$ is $\Theta(k\theta\log n)$ as the sum of $n$ independent variables. Therefore, by a standard Chernoff bound as in \Cref{lem:chernoff}, we can obtain this result.
\end{proof}

Then we show that, with high probability, for all $|C|>n/\theta$ where $\delta^-(C)$ is a $s$-rooted minimal cut, there exists some sampled $q_i \in C$. We note that in this case, $\delta^-(C)$ is also an $(s,t)$ minimal cut which is also in-reachable, for all $t\in C$. In fact, this property holds for every in-reachable cut, as we show below.

\begin{property}
    \label{ass:smallcuts}
    For each vertex $t$ and each in-reachable $(s,t)$ cut $(S,T)$ with $\rho(T) \leq k$ and $|T| > n/\theta$, we have $T \cap Q \neq \emptyset$.
\end{property}

\begin{lemma}
    \Cref{ass:smallcuts} holds with high probability.
\end{lemma}

\begin{proof}
    We first notice that, given a vertex $t$ and the $(s,t)$ cut edges $Z$ with $|Z|\leq k$, one can uniquely define $T$ as all vertices that can reach $t$. Since we have $O(n)$ choices of $t$ and $O(m^k)$ choices of $Z$, there are at most $O(nm^k) = O(n^{2k+1})$ possible such sets $T$.

    For each $T$, we know $\Pr[T \cap Q = \emptyset] = (1-\Theta(k\theta\log n/n)^{(n/\theta)}) = n^{-\Theta(k)}$, therefore, by a union bound on all the $O(n^{2k+1})$ possible such sets $T$, the lemma holds.
\end{proof}

\begin{remark}
    For simplicity, we will assume that \Cref{ass:samplesize} and \Cref{ass:smallcuts} hold through all this section, since they all hold with high probability and are both verifiable in polynomial time. Therefore, even if either property fails throughout our augmentation, one can simply sample again to guarantee both of them will hold.
\end{remark}

Based on this, we know for all $|T|>n/\theta$ where $G[T]$ is strongly connected and $\delta^-(T)$ is a $s$-rooted minimal cut, there exists some sampled $q \in T$. We show that, by using an $(s,q)$ flow-augmentation from \Cref{thm:staugment}, $C$ will become an $s$-rooted min-cut.

\begin{lemma}
    \label{lem:augment-sq}
    Suppose $\delta^-(T)$ is an $s$-rooted minimal cut with capacity at most $k$ and $q \in T$. Let $A$ be a random arc set generated by \Cref{thm:staugment} for $(s,q)$ cuts, and let $B_q = \{(q,v)|v \in V-\{s\}\}$, then with probability $2^{-O(k^4\log k)}$, $\delta^-(T)$ becomes $s$-rooted min-cut in $G+A+B_q$.
\end{lemma}

\begin{proof}
    We first observe that $\delta^-(T)$ is also an $(s,q)$ minimal cut for any $q \in T$, since $s$ can reach $q$ in $G-\delta^-(T)$ but not for any proper subset of $\delta^-(T)$. Therefore, from \Cref{thm:staugment}, with probability $2^{-O(k^4\log k)}$, $\delta^-(T)$ becomes an $(s,q)$ min-cut in $G+A$. Now we consider $G+A+B_q$. Clearly, $\delta^-(T)$ remains an $(s,q)$ min-cut. Suppose $(S',T')$ is an $s$-rooted min-cut, then $q \in T'$ must hold since otherwise $\delta^-(T')$ contains $B_q$ edges from $q$ to every vertex in $T'$ (with $+\infty$ capacity). Therefore, any $s$-rooted min-cut must be an $(s,q)$ min-cut as well and we know $\delta^-(T)$ becomes an $s$-rooted min-cut.
\end{proof}

Now we focus on all small components $T$ with $T \cap Q= \emptyset$. We define an edge set $A_Q = \{(s,q)|q \in Q\}$ and consider the graph $G+A_Q$ (recall that edges in $A_Q$ have capacity $+\infty$, so this is equivalent to contracting $Q$ into $s$), in which \Cref{ass:smallcuts} implies the following observation:

\begin{lemma}\label{lem:smallcuts}
    Suppose \Cref{ass:smallcuts} holds, then in graph $G+A_Q$, for each vertex $t$ and each in-reachable $(s,t)$ cut $(S,T)$ with $\rho_{G+A_Q}(T) \leq k$, we have $|T| \leq n/\theta$.
\end{lemma}

\begin{proof}
    Suppose there exists $|T|>n/\theta$ that $\rho_{G+A_Q}(T) \leq k$, by \Cref{ass:smallcuts}, $T \cap Q \neq \emptyset$, therefore, there exists some $q \in Q \cap T$ and $(s,q) \in A_q$. Therefore, $\delta^-_{G+A_Q}(T)$ contains edge $(s,q)$ with $+\infty$ capacity and this contradicts $\rho_{G+A_Q}(T) \leq k$. Therefore, the above observation holds.
\end{proof}

Suppose the condition of the above lemma holds. Then, in $G+A_Q$, for every $t$ and every in-reachable $(s,t)$ cut $(S,T)$, we know that $T$ has relatively small size. Next, from \Cref{lem:impsep}, for each vertex $t$, we can define a family $\{T_t^{(i)}\}$ containing all $(s,t)$ important separators $(V - T_t^{(i)}, T_t^{(i)})$, where $T_t^{(i)}$ is the $i$-th important separator. From \Cref{lem:impsep}, there are at most $4^k$ important separators (so $i \in [4^k]$). Based on \Cref{ass:smallcuts}, each important separator has size $|T_t^{(i)}| \leq n/\theta$.

Then for each important separator $T_t^{(i)}$, we define $C_t^{(i)} \subseteq T_t^{(i)}$ as the strongly connected component in $G[T_t^{(i)}]$ containing $t$ (we note that $\rho(C_t^{(i)})$ may be larger than $k$). We have $|C_t^{(i)}| \leq n/\theta$. Let $\mathcal{C} = \{C_t^{(i)}| \not\exists t',j \text{ s.t. } C_t^{(i)} \subseteq C_{t'}^{(j)}\}$ be the distinct collection of all (inclusion-wise) maximal components $C_t^{(i)}$ for all $t$ and $i$. Next, we show some properties of the family $\mathcal{C}$. We first state a simple observation about important separators that will be used below.

\begin{lemma}\label{lem:impsep-eq}
    Suppose $(S,T)$ is an important separator for vertices $s,t \in G$, and $t$ and $t'$ are strongly connected in $G[T]$, then $(S,T)$ is an important separator for vertices $s$ and $t'$.
\end{lemma}

\begin{proof}
    Since $t$ and $t'$ are strongly connected in $G[T]$ and every vertex $v \in T$ can reach $t$ in $G[T]$, we know every vertex $v \in T$ can also reach $t'$ in $G[T]$. Thus, $(S,T)$ is a in-reachable cut for $s$ and $t'$ as well. Now suppose towards contradiction that there is some other in-reachable cut $(S',T')$ for $t'$ with $T \subseteq T'$, we observe that $t$ and $t'$ are also strongly connected in $G[T']$ and thus $(S',T')$ is also a in-reachable cut for $t$. But $T \subseteq T'$ and $T$ is an important separator for $(s,t)$, so such $T'$ does not exist and thus $T$ is an important separator for $s$ and $t'$.
\end{proof}

\begin{lemma}
    \label{lem:impscc-t}
    For any $C \in \mathcal{C}$ and any vertex $t \in C$, $C=C_t^{(i)}$ for some $i$.
\end{lemma}

\begin{proof}
    The lemma comes directly from \Cref{lem:impsep-eq}. Since $C \in \mathcal{C}$, it must come from some $v \in C$ that $C = C_{v}^{(j)}$. Therefore, for $T = T_{v}^{(j)}$, since $t$ and $v$ are strongly connected in $G[T]$, we know $(S,T)$ is also an important separator for $s$ and $t$ by \Cref{lem:impsep-eq}. Since $C$ is also the strongly connected component containing $t$ inside the $(s,t)$ important separator $(S,T)$, we know $C=C_t^{(i)}$ for some $i$.
\end{proof}

\begin{lemma}
    \label{lem:impscc-family}
    Suppose \Cref{ass:smallcuts} holds, in the graph $G+A_Q$, the family $\mathcal{C}$ has the following properties:
    \begin{enumerate}
        \item $\forall C \in \mathcal{C}$, $|C|\leq n/\theta$.
        \item $\sum_{C \in \mathcal{C}} |C| \leq n \cdot 4^k$.
        \item For any $s$-rooted minimal cut $Z = \rho^-(T)$ where $T$ is strongly connected, we have $T \subseteq C$ for some $C \subseteq \mathcal{C}$.
    \end{enumerate}
\end{lemma}

\begin{proof}
    We prove these properties as follows.
    \begin{enumerate}
        \item Suppose C = $C_t^{(i)}$ for some $t \in C$ and $i$. Let $T = T_{t}^{(i)}$. Since $T$ is an $(s,t)$ in-reachable cut in $G+A_Q$, from \Cref{lem:smallcuts}, we know $|T| \leq n/\theta$. Therefore, since $C \subseteq T$, we know $|C| \leq n/\theta$.
        \item For each vertex $v$, from \Cref{lem:impsep}, there are at most $4^k$ important separators $T_t^{(i)}$ and thus it is corresponding to at most $4^k$ components $C_t^{(i)}$. We count all $4^kn$ pairs $(t,i)$ with $t \in V$ and $i \leq 4^k$. By \Cref{lem:impscc-t}, for each component $C\in \mathcal{C}$, we know $C$ is corresponding to some pair $(t,i)$ for each $t \in C$. Therefore, it is corresponding to $|C|$ pairs in total. Since there are at most $4^kn$ pairs, we have $\sum_{C \in \mathcal{C}} |C| \leq n \cdot 4^k$.
        \item For any $s$-rooted minimal cut $Z = \rho^-(T)$ where $T$ is strongly connected, we pick an arbitrary $t \in T$. By the maximality of important separators $T_t^{(i)}$, there exists an $i$ that $T \subseteq T_t^{(i)}$. Since $T$ is also strongly connected, it must be fully contained in $C_{t}^{(i)}$, since it is the strongly connected component in a larger subgraph $G[T_t^{(i)}]$. Finally, by the maximality of components $C \subseteq \mathcal{C}$, there exists some $C \subseteq \mathcal{C}$ that $C_{t}^{(i)} \subseteq C$, therefore, we have $T \subseteq C_{t}^{(i)} \subseteq C$.
    \end{enumerate}
\end{proof}

To partition and recurse in a next level, we show the following lemmas holds:

\begin{lemma}
    \label{lem:impscc-partition}
    Given \Cref{ass:smallcuts} and the family $\mathcal{C}$ as in \Cref{lem:impscc-family}, we can partition $\mathcal{C}$ into $O(k4^k)$ subfamilies $\mathcal{C}_i$, where each $\mathcal{C}_i$ forms a subpartition of $V$ (i.e., all $C \in \mathcal{C}_i$ are disjoint vertex subsets in $V$).
\end{lemma}

\begin{proof}
    Consider any $C \in \mathcal{C}$. From \Cref{lem:impscc-t}, we suppose $C = C_t^{(i)}$ for some $t$ and let $T = T_t^{(i)}$ be the corresponding important separator containing $C$. Now suppose some different $C' \in \mathcal{C}$ has $C \cap C' \neq \emptyset$. We observe that $C' \not\subseteq C$ since all components are inclusion-wise maximal in $\mathcal{C}$. Suppose $C' \subseteq T$, then $C \cup C'$ is a larger strongly connected set inside $G[T]$, which contradicts the fact that $C$ is a strongly connected component in $G[T]$. Therefore, by contradiction we have $C' \not \subseteq T$.
    
    Since $G[C']$ is strongly connected and $C'$ contain vertices both inside and outside $T$, we know $C' \cap N^-(T) \neq \emptyset$. Let $v \in C' \cap N^-(T)$, then $C' = C_v^{(i)}$ for some $i$ from \Cref{lem:impscc-t}. We observe that $|N^-(T)| \leq \rho(T) \leq k$ and each vertex $v \in N^-(T)$ can have at most $4^k$ different $C' = C_v^{(i)}$ by \Cref{lem:impsep} and the definition of $C_v^{(i)}$. Therefore, each $C \in \mathcal{C}$ intersects at most $k4^k$ different $C' \in \mathcal{C}$.

    Consider an undirected auxiliary graph $G_\mathcal{C}$ with vertex set $\mathcal{C}$ and there is an edge $(C,C')$ if $C \cap C' \neq \emptyset$. By above we have the maximum degree in $G_\mathcal{C}$ is at most $k4^k$ and thus $G_\mathcal{C}$ is $(k4^k+1)$-colorable. For each of the $(k4^k+1)$ colors, we construct a subfamily $\mathcal{C}_i$. Therefore, for any $C,C' \in \mathcal{C}_i$, we have $(C,C') \notin G_\mathcal{C}$ and thus $C \cap C' \neq \emptyset$ and thus the $(k4^k+1)$ subfamilies $\mathcal{C}_i$ we construct will satisfy the statement.
\end{proof}

\subsubsection{Dealing with Multiple Levels}

Below, we aim to prove \Cref{thm:weak-augment} recursively using \Cref{alg:weak-augment}.

\begin{algorithm}[ht]

\caption{Many-Graph Version of Rooted Flow-Augmentation}
\begin{algorithmic}[1] 
\label{alg:weak-augment}

\Procedure{RootedFlowAugmentation}{$G=(V,E), \theta, \ell$} \Comment{$\theta = 2^{\sqrt{k \log n}}, \ell \gets [0,L]$}

    \State Randomly sample $Q$ as in \Cref{ass:samplesize}
    \If{$\ell = 0$}
        \State Uniformly randomly sample a vertex $q \gets Q$
        \State Apply \Cref{thm:staugment} for $s$ and $q$ to get a random edge set $A$
        \State $B_q = \{(q,v)|v \in V-\{s\}\}$
        \State \Return $\mathcal{R} = \{V-\{s\}\}$, $\mathcal{T} = \{q\}$, and $\mathcal{A} = \{A \cup B_q\}$
    \Else
        \State Compute $\mathcal{C}$ in graph $G+A_Q$
        \State Uniformly randomly get one $\mathcal{C}_i$ in subfamilies $\{\mathcal{C}_i\}$ obtained from \Cref{lem:impscc-partition}
        \For{$C \in \mathcal{C}_i$}
            \State Let $H$ be the graph obtained from $G$ by contracting all vertices in $V-C$ to $s$
            \State \Call{RootedFlowAugmentation}{$H,\theta,\ell-1$}
        \EndFor
        \State \Return $\mathcal{R},\mathcal{T},\mathcal{A}$ as the union of returned $\mathcal{R},\mathcal{T},\mathcal{A}$ for all $H$
    \EndIf
\EndProcedure

\end{algorithmic}
\end{algorithm}

\begin{remark}
    \label{rem:ell}
    We remark that, the procedure in \Cref{alg:weak-augment} is identical (in distribution of returned sets $\mathcal{R},\mathcal{T},\mathcal{A}$) to the following scheme:
    \begin{itemize}
        \item First, compute and random choose a subpartition $\mathcal{C}_i$ in each level and recurse. Since the instance size in each level will be reduced by a factor of $\theta$ due to \Cref{ass:smallcuts}, there are at most $L = \log_\theta n$ levels.
        \item Then, choose a random $\ell \in [0,L]$, a random sample $q$ in recursion level $\ell$ and perform flow-augmentations to get $G_i+A_i+B_q$ for all graphs $G_i$ in the $\ell$-th layer of recursion.
    \end{itemize}
\end{remark}

The brief proof strategy goes as follows. Let's first suppose the case that $(S,T)$ is an $s$-rooted minimal cut that when we sample $Q$ from the whole graph in our first level of recursion, there exists some $q \in Q\cap T$. Therefore, a way to capture $(S,T)$ is by a flow-augmentation for $(s,q)$. The probability that we make a flow-augmentation for $(s,q)$ is based on $\ell=0$ (with probability $\Omega(1/L)$), we sample the correct vertex $q \in Q$ (with probability $1/|Q|$), and the cut becomes an $(s,q)$ min-cut in the flow-augmentation for $(s,q)$ (with probability $2^{-\poly(k)}$).

When $(S,T)$ is not captured by any $q \in Q\cap T$ in the first level, here, by \Cref{lem:impscc-family}, we know $T \subseteq C$ for some $C \in \mathcal{C}$, and the good event happens when we samples the subfamily $\mathcal{C}_i \in C$ to recurse. To analyze the probability, we note that there are $O(k4^k)$ such families in each level of recursion, and the probability may multiply through all at most $L$ layers. This gives another $(k4^k)^{-O(L)}$ factor to the final success probability. We prove \Cref{thm:weak-augment} formally below.

\begin{proof}[Proof of \Cref{thm:weak-augment}]
    We consider the procedure in \Cref{alg:weak-augment}. Let $\theta = 2^{\sqrt{k\log n}}$ and $L = \log_\theta n = \sqrt{\log n/k}$. We first randomly choose a level $\ell$ from $0$ to $L$. Here we assume both \Cref{ass:samplesize} and \Cref{ass:smallcuts} hold in all our recursions. Since in any recursion, they happen with high probability and there are $O(n)$ instances in total that we recurse on, we know all of them will hold simultaneously with high probability.
    
    It is clear that we always return $\mathcal{R}$ as a subpartition of $V-\{s\}$ and $\mathcal{A}$ accordingly that $A_i \subseteq (R_i \cup \{s\})\times R_i$ since we always recurse in one $\mathcal{C}_i$ that contains disjoint vertex subsets $C$. Suppose $(S,T)$ is a $s$-rooted minimal cut with size at most $k$ where $T$ is strongly connected. Given the random sample $Q$, there are two possible cases: $T \cap Q \neq \emptyset$; and $T \cap Q = \emptyset$.
    
    Conditioning on the first case that $T \cap Q \neq \emptyset$. Suppose now $\ell = 0$ and we are performing a flow-augmentation to get $A+B_q$. Then, based on \Cref{ass:samplesize}, since $|Q| = O(k\theta\log n)$ and $T \cap Q \neq \emptyset$, we sample some $q$ inside $T$ successfully with probability $\Omega((k\theta\log n)^{-1})$. Finally, given that some $q\in T$ is sampled, based on \Cref{lem:augment-sq}, with probability $2^{-O(k^4\log k)}$, $\delta^-(T)$ becomes an $s$-rooted mincut in $G+A+B_q$.

    Now suppose the second case that $T \cap Q = \emptyset$. Here, by \Cref{lem:impscc-family}, we know $T \subseteq C$ for some $C \in \mathcal{C}$. Since we partition $\mathcal{C}$ into $O(k4^k)$ subfamilies $\mathcal{C}_i$, we will sample a $\mathcal{C}_i \ni C$ with probability $\Omega(1/k4^k)$. In the next steps, we contract $V-C$ into $s$ to form graph $H$. Note that $\delta_H^-(T)$ is still an $s$-rooted minimal cut in $H$, so when we recurse in $H$, the same analysis as above is still valid.

    Given \Cref{ass:smallcuts}, from \Cref{lem:impscc-family}, each time we recurse into smaller graphs $H$ from $G$, the number of vertices is divided by $\theta$. Therefore, the recursion can run for at most $L = \log_\theta n$ layers. Consider the good event that, in all layers before we sample some $q \in T$, $T \subseteq C$ always holds for some $C \in \mathcal{C}_i$ for the family $\mathcal{C}_i$ we choose along recursions. Since each layer the probability is $\Omega(1/k4^k)$, and there are at most $L$ preceding layers, the overall probability for this good event has an lower bound $(k4^k)^{-O(L)}$.

    When the good event occurs, we also need from \Cref{rem:ell} that a correct $\ell \in [0,L]$ is sampled, which is with probability $\Omega(1/L)$. From above we know $\delta^-(T)$ becomes a $s$-rooted mincut in $H+A+B_q$ with probability $2^{-O(k^4\log k)} \times \Omega((k\theta\log n)^{-1})$.
    
    Therefore, the overall probability that $\delta^-(T)$ become an $s$-rooted minimum cut in $H+A+B_q$ will be \[(k4^k)^{-O(L)} \times L^{-1} \times 2^{-O(k^4\log k)} \times \Omega((k\theta\log n)^{-1}) = 2^{-O(\sqrt{k\log n} + k^4\log k)},\]
    when we choose $L = \sqrt{\log n/k}$ and $\theta = n^{1/L} = 2^{\sqrt{k\log n}}$.
\end{proof}


\begin{remark}
    By a standard argument, the term $2^{-O(\sqrt{k\log n})}$ is at least $2^{-\poly(k)} n^{-o(1)}$, since if $k \leq \sqrt{\log n}$, it is at least $n^{-o(1)}$, and otherwise, it is at least $2^{-\poly(k)}$.
\end{remark}


\subsection{Flow-Augmentation}

In this section, we present the full proof for \Cref{thm:strong-augment} by exploiting more properties for the family $\mathcal{C}$ containing maximal SCCs as in \Cref{lem:impscc-family}. 

\strongaugment*

We first show a stronger and more careful partition of $\mathcal{C}$ in \Cref{lem:impscc-family} into subfamilies $\mathcal{C}_i$ than \Cref{lem:impscc-partition}. Intuitively, instead of merely claiming any different subsets $C,C' \in \mathcal{C}_i$ are vertex disjoint in each subfamily, we now want to claim that $C,C' \in \mathcal{C}_i$ are not only disjoint, but also independent in the sense that they are not linked to each other by any edge. This is crucial if we want to integrate the many-graph version to the original graph later.

To obtain the stronger condition, we need to look into the structure of all sets $C \in \mathcal{C}_i$ (obtained by \Cref{lem:impscc-partition}). The strategy goes as follows: First, we define a canonical important separator $T_C$ for each $C \in \mathcal{C}$, then, based on the relation for important separators $T_C$, we can define a partial order on all sets $C$ in \Cref{lem:partialorder}. Finally, we use a decomposition of DAGs in \Cref{lem:DAGdecomp} to partition all $C \in \mathcal{C}_i$ into subfamilies as mentioned above.

Below we still consider the graph $G+A_Q$ as stated in \Cref{lem:smallcuts} and the family $\mathcal{C}$ as in \Cref{lem:impscc-family}. Instead of getting all components $C$ with an implicit separator $(S,T)$, we first define a canonical important separator that is corresponding to each $C \in \mathcal{C}$, which can be computed given the component $C$ itself:


\begin{lemma}
    Let $C \in \mathcal{C}$ and let $(S,T)$ be the furthest in-reachable $(s,C)$ min-cut. Then for any $t \in C$, $(S,T)$ is an $(s,t)$ important separator.
\end{lemma}

\begin{proof}
    Suppose towards contradiction that $(S,T)$ is not a $(s,t)$ important separator, then there must exist another \emph{in-reachable} $(s,t)$ cut $(S',T')$ where $T \subseteq T'$ with $\rho(T') \leq \rho(T)$ (recall that $\rho(T)$ is defined as the capacity of $\delta^-(T)$). We can also observe that $(S',T')$ is an $(s,C)$ \emph{in-reachable} cut, thus, it must be an $(s,C)$ in-reachable \emph{min-cut}. However, since $T \subseteq T'$, it contradicts the fact that $(S,T)$ be the furthest in-reachable $(s,C)$ min-cut. Therefore, the lemma holds\footnote{The lemma can also be derived from \Cref{lem:impsep-eq} by (informally) treat $C$ as a sink.}.
\end{proof}

Below, for each $C \in \mathcal{C}$, we use $(S_C,T_C)$ to denote the furthest $(s,C)$ in-reachable min-cut. Our approach first apply a similar partitioning to \Cref{lem:impscc-partition}, and then look into each subfamily $\mathcal{C}_i$ and partition a little further.

\begin{lemma}
    $\mathcal{C}$ can be partitioned into $O(k^24^k)$ subfamilies $\{\mathcal{C}_i\}$ such that in each family $\mathcal{C}_i$:
    \begin{itemize}
        \item For any $C,C' \in \mathcal{C}_i$, $C' \cap N^-(T_C) = \emptyset$.
        \item For any $C,C' \in \mathcal{C}_i$, $\rho(T_C) = \rho(T_{C'})$.
    \end{itemize}
\end{lemma}

\begin{proof}
    The proof is a little stronger than \Cref{lem:impscc-partition}. We build an auxiliary directed graph $H$ with vertex set $\mathcal{C}$ that an edge $(C,C') \in H$ if $C' \cap N^-(T_C) \neq \emptyset$. Since $|N^-(T_C)| \leq \rho(T_C) \leq k$, and each $v \in N^-(T_C)$ is inside at most $4^k$ different $C'$ components, we know the out-degree for each vertex $C \in \mathcal{C}$ is at most $k4^k$. Therefore, if we treat $H$ as an undirected graph, we know $H$ is $2k4^k$-degenerate and thus $H$ is $(2k4^k+1)$-colorable. Therefore, we can partition $\mathcal{C}$ into $O(k4^k)$ subfamilies $\mathcal{C}_i$ based on colors in $H$. Then for any $C,C'$ in the same $\mathcal{C}_i$, $C' \cap N^-(T_C) = \emptyset$ since $(C,C') \notin H$.

    Therefore, the partition satisfies the first property. To also satisfy the second property, since $\rho(T_C) \leq k$ for any $C \in \mathcal{C}$, we can further partition each $\mathcal{C}_i$ into $k$ families based on $\rho(T_C) \leq k$, that is, for each $k' \leq k$, we put all $C \in \mathcal{C}_i$ with $\rho(T_C)=k'$ into one family. Therefore, after the above further partition, we have a multiplication by $k$ and there are $O(k^24^k)$ subfamilies in total, and each subfamily $\mathcal{C}_i$ will satisfy both properties.
\end{proof}

We note that same as in \Cref{lem:impscc-partition}, for any strongly connected set $C' \cap N^-(T_C) = \emptyset$, we must have either $C' \cap T_C = \emptyset$ or $C' \subseteq T_C$ (in this case we still have $C' \cap C = \emptyset$ since otherwise $C \cup C'$ forms a larger strongly connected set). 

Now let's consider both two cases. In the first case that $C' \cap T_C = \emptyset$, since $C' \cap N^-(T_C) = \emptyset$ also holds, we know there is no edge from $C'$ to $C$. In the second case, there may still be edges from $C'$ to $C$. However, in this case we have the following observation:

\begin{lemma}
    \label{lem:partialorder}
    Inside $\mathcal{C}_i$, suppose there is an edge from $C'$ to $C$ (so $C' \subseteq T_C - C$) and $\rho(T_C) = \rho(T_{C'})$, then $T_{C'} \subset T_{C}$ is a proper subset of $T_C$.
\end{lemma}

\begin{proof}
    We first prove $T_{C'} \subseteq T_{C}$. By submodularity we have \[2 \rho(T_{C'}) = \rho(T_C) + \rho(T_{C'}) \geq \rho(T_C \cap T_{C'}) + \rho(T_C \cup T_{C'}).\]
    Since the right side are still corresponding to two $(s,C')$ cuts, we know $\rho(T_C \cap T_{C'}) + \rho(T_C \cup T_{C'}) \geq 2\rho(T_{C'})$, so $\rho(T_C \cap T_{C'}) = \rho(T_C \cup T_{C'}) = \rho(T_{C'}) = \rho(T_{C})$. Therefore we know $\delta^-(T_C \cup T_{C'})$ is also a $(s,C)$ min-cut. Moreover, it is in-reachable from $C$ since $T_C$ is in-reachable from $C$ and $T_{C'}$ is in-reachable from $C' \subseteq T_C$. Therefore, by the maximality of $T_C$, we can only have $T_C \cup T_{C'} \subseteq T_C$ and thus $T_{C'} \subseteq T_{C}$.

    Then we prove $T_{C'} \neq T_{C}$. Suppose that $T_{C'} = T_{C}$, since they are both in-reachable cuts, we know $C'$ and $C$ are reachable to each other in $G[T_C]$, so they must be strongly connected in $G[T_C]$. This is a contradiction and thus we have $T_{C'} \subset T_{C}$.
\end{proof}

Therefore, one can build a directed auxiliary graph $H$ with vertex set $\mathcal{C}_i$ and $(C',C) \in H$ if there is an edge from $C'$ to $C$ in the original graph. We observe that, by the above lemma, if $(C',C) \in H$ then $T_{C'} \subset T_{C}$. Therefore, base on the inclusion-wise partial order of the $T_C$ sets, we deduce that $H$ is a DAG. Now, we use the following decomposition of $H$ to further partition $\mathcal{C}_i$ into smaller families.


\paragraph{A Decomposition of DAGs.} We consider the following decomposition of $H$.

\begin{lemma}
    \label{lem:DAGdecomp}
    Let $H=(V,E)$ be a DAG, then there exist disjoint vertex sets $\{S_i\}$, such that:
    \begin{itemize}
        \item For any $i \neq j$, there is no edge between $S_i$ and $S_j$ in each direction.
        \item For any $i$, there exists a vertex $t_i$ such that every vertex in $S_i$ can reach $t_i$ in $H$.
        \item $\sum_i|S_i| = \Omega(n/\log n)$.
    \end{itemize}
\end{lemma}

\begin{proof}
    Let $T = \{t \in V| \delta^+_H(t)=\emptyset\}$ be the set of \emph{sinks} in $H$. For each vertex $v \in V$, we define $R_v$ to be the set of vertices in $T$ reachable by $v$. When there's an edge $(u,v)$, we must have $R_v \subset R_u$. Fix any $I \subseteq T$, for any $t \in I$, we let $S_t = \{v\ \in V|R_v \cap I = \{t\}\}$ be the set of vertices that can only reach $t$ but not any other vertex in the subset $I$. Therefore, suppose there are two vertices $t,t' \in I$ and $v \in S_t, v' \in S_{t'}$, there cannot be an edge $(v,v')$ since otherwise $v$ can also reach $t'$.

    Now we aim to choose a good subset $I \subseteq V$. We can partition $V$ into $\log n$ subsets $V_i = \{v \in V| \space |R_v| \in [2^i,2^{i+1})\}$. By pigeonhole principle, there exists some $i$ that $|V_i| \geq n/\log n$. We choose each vertex in $T$ with a uniform probability $2^{-i}$ to form the subset $I$. Now, for each $v \in V_i$, \[\Pr[|R_v \cap I|=1] = |R_v| \cdot 2^{-i} \cdot (1-2^{-i})^{|R_v|-1} = \Omega(1)\]
    since $|R_v| \in [2^i,2^{i+1})$.

    Therefore, in expectation, $\Omega(n/\log n)$ vertices in $V_i$ are inside some $S_t$ where $t \in I$. That is saying, $\sum_{t \in I} |S_t| = \Omega(n/ \log n)$. Therefore, by picking $\{S_t\}$ we know the lemma holds.
\end{proof}

\begin{lemma}
    \label{lem:impscc-partition-full}
    For any $\mathcal{C}_i$, there exist $O(\log^2n)$ families $\mathcal{D}_{ij}$ such that:
    \begin{itemize}
        \item For each vertex set $D$ in $\mathcal{D}_{ij}$, it is the disjoint union of some sets $C \in \mathcal{C}_i$, and there exists some $C \in \mathcal{C}_i$ where $D \subseteq T_C$ (and hence $|D| \leq n/\theta$).
        \item For each $C \in \mathcal{C}_i$, it is subset of exactly one $D$ in exactly one $\mathcal{D}_{ij}$.
        \item For any $D,D' \in \mathcal{D}_{ij}$ in the same subfamily, there's no edge between $D$ and $D'$.
    \end{itemize}
    Therefore, there exist $O(k^24^k \log^2 n)$ subfamilies $\mathcal{D}_{ij}$ of $\mathcal{C}$ in total.
\end{lemma}

\begin{proof}
    Consider the auxiliary DAG $H$ with vertex set $\mathcal{C}_i$ and $(C',C) \in H$ if there is an edge from $C'$ to $C$ in the original graph that we built above. By \Cref{lem:DAGdecomp}, we obtain disjoint vertex sets $\{S_i\}$ and let $D_i \subseteq V$ be the union of all $C \in \mathcal{C}_i$ that is contained in $S_i$. This will give us one family $\mathcal{D} = \{D_i\}$. By \Cref{lem:DAGdecomp}, there exists some $C \in \mathcal{C}_i$ that all components in $S_i$ can reach $C$. Therefore, by \Cref{lem:partialorder}, all components in $S_i$ are inside $T_C$, so we obtain $D \subseteq T_C$. One can also see that, if there is an edge between two different vertex subsets $D_i,D_j \in \mathcal{D}$, then there should be an edge between $S_i$ and $S_j$ in $H$, which contradicts \Cref{lem:DAGdecomp}.

    Finally, we prove the total number of families $\mathcal{D}_{ij}$. Since we can choose a family $\mathcal{D}$ each time that corresponds to $\Omega(|\mathcal{C}_i|/\log|\mathcal{C}_i|)$ components in $\mathcal{C}_i$, we know by repeatedly choosing such $\mathcal{D}$ and delete all $\Omega(|\mathcal{C}_i|/\log|\mathcal{C}_i|)$ components, with $O(\log^2|\mathcal{C}_i|)$ rounds we will make $H$ empty. Therefore, these $O(\log^2 n)$ families will satisfy the lemma statement.
\end{proof}

Now given the partition of every $\mathcal{C}_i$ into $\mathcal{D}_{ij}$, we first want to modify \Cref{alg:weak-augment} by recursing on all subfamilies $\mathcal{D}_{ij}$ instead of all $\mathcal{C}_i$. Moreover, after returning the final subpartition $\mathcal{R}$ and edge sets $\mathcal{A}$, we need to assemble the whole graph by adding some edge set $A$ in $G$ to make the $s$-rooted min-cuts in $G_i$ still $s$-rooted min-cuts in $G$ for every contracted graph $G_i$ simultaneously. In fact, this requires the property that $\mathcal{R}$ is not only disjoint but also independent. 

Below, we will show an analog of \Cref{alg:weak-augment} to give such a subpartition $\mathcal{R}$, and explain how to assemble into the whole graph, given the subpartition $\mathcal{R}$ and edge sets $\mathcal{A}$. We first state a extended version of \Cref{thm:weak-augment} as below, then we modify \Cref{alg:weak-augment} into \Cref{alg:full-augment} and assemble the outcomes of \Cref{alg:full-augment} in the whole graph $G$. For technical reasons, here we state the version when the minimal cut is with size \emph{exactly} $k$ instead of \emph{at most} $k$. The main modified parts are colored in blue.

\begin{algorithm}[ht]

\caption{Rooted Flow-Augmentation}
\begin{algorithmic}[1] 
\label{alg:full-augment}

\Procedure{RootedFlowAugmentation}{$G=(V,E), \theta, \ell$} \Comment{$\theta = 2^{\sqrt{k \log n}}, \ell \gets [0,L]$}

    \State Randomly sample $Q$ as in \Cref{ass:samplesize}
    \If{$\ell = 0$}
        \State Uniformly randomly sample a vertex $q \gets Q$
        \State Apply \Cref{thm:staugment} for $s$ and $q$ to get a random edge set $A$
        \State $B_q = \{(q,v)|v \in V-\{s\}\}$
        \If{{\color{blue}$G+A+B_q$ is not $s$-rooted $k$-connected}}
            \State {\color{blue} Add $(s,q)$ to $B_q$}
        \EndIf
        \State \Return $\mathcal{R} = \{V-\{s\}\}$, $\mathcal{T} = \{q\}$, and $\mathcal{A} = \{A \cup B_q\}$
    \Else
        \State Compute $\mathcal{C}$ in graph $G+A_Q$
        \State {\color{blue}Uniformly randomly get one $\mathcal{D}$ in subfamilies $\{\mathcal{D}_{ij}\}$ obtained from \Cref{lem:impscc-partition-full}}
        \For{$D \in \mathcal{D}$}
            \State Let $H$ be the graph obtained from $G$ by contracting all vertices in $V-D$ to $s$
            \State \Call{RootedFlowAugmentation}{$H,\theta,\ell-1$}
        \EndFor
        \State \Return $\mathcal{R},\mathcal{T},\mathcal{A}$ as the union of returned $\mathcal{R},\mathcal{T},\mathcal{A}$ for all $H$
    \EndIf
\EndProcedure

\end{algorithmic}
\end{algorithm}

Below, we still assume both \Cref{ass:samplesize} and \Cref{ass:smallcuts} hold in all our recursions, since all of them will hold simultaneously with high probability same as in \Cref{alg:weak-augment}. We will list some observations about \Cref{alg:full-augment} below and use the returned families in \Cref{alg:full-augment} to finally prove \Cref{thm:strong-augment}.

\begin{lemma}
    \Cref{alg:full-augment} will return a family $\mathcal{R}$ such that for any $R_i,R_j \in R$, there is no edge between $R_i$ and $R_j$.
\end{lemma}

\begin{proof}
    The lemma comes directly from our algorithm. If $\ell = 0$, then $\mathcal{R}$ is a singleton and the lemma holds trivially. In the case $\ell > 0$, since we recurse in a vertex subset family $\mathcal{D}$ that satisfies \Cref{lem:impscc-partition-full}, we know for each different $D,D'$ we recurse on, there is no edge between $D$ and $D'$. Therefore, by a simple induction on $\ell$ we will obtain this lemma.
\end{proof}

\begin{lemma}
    \label{lem:full-anal1}
    For any $s$-rooted minimal cut $(S,T)$ with capacity $\leq k$ and $T$ is strongly connected, The following will happen simultaneously with probability $(1/L) \cdot (k^24^k\log^2 n)^{-O(L)}$:
    \begin{itemize}
        \item There exists $R \in \mathcal{R}$ that $T \subseteq R$.
        \item There is a $q \in T$ sampled when $\ell$ becomes $0$ and we recurses in the above $R$.
    \end{itemize}
\end{lemma}

\begin{proof}
    We note in the same way as \Cref{rem:ell} that, the algorithm can be seen as getting subfamily $\mathcal{D}$ in all (at most $L$) layers first, then sample $\ell$ and return all subsets in layer $\ell$ as $\mathcal{R}$. For each $T$, from \Cref{lem:impscc-family} and \Cref{lem:impscc-partition-full}, we know in each layer, as long as we do not sample a $q \in T$, there exists some $\mathcal{D}$ in all $O(k^24^k\log^2n)$ subfamilies that some $D \in \mathcal{D}$ will include $T$ as a subset. Therefore, the probability that we always find such a correct $\mathcal{D}$ is $O(1/k^24^k\log^2n)$ in each layer, and at least $(k^24^k\log^2n)^{-O(L)}$ in all layers before we sample some $q \in T$ at some level $x$ in the recursion. Finally, we just need the initially sampled $\ell$ equals $x$, and this is with probability $1/L$.
\end{proof}

\begin{lemma}
    \label{lem:full-anal2}
    Given the conditions in the above lemma, we let $G_i$ be the graph obtained by contracting $V-R_i$ to the source $s$ in $G$. Then for any $s$-rooted minimal cut $(S,T)$ in $G_i$ with capacity \textbf{exactly} $k$ and $T$ is strongly connected, $(S,T)$ becomes an $s$-rooted min-cut inside $G_i+A+B_q$ with probability $|Q|^{-1} \cdot 2^{-O(k^4\log k)}$.
\end{lemma}

\begin{proof}
    We consider the branch when $\ell = 0$ and there is some $q \in T$ sampled. Since we choose a random $q \in Q$, the probability that we choose a $q \in T$ will be $|Q|^{-1}$. Next, suppose a correct $q \in T$ is sampled, from \Cref{lem:augment-sq}, with probability $2^{-O(k^4\log k)}$ $\delta^-(T)$ will become an $(s,q)$ min-cut and also and $s$-rooted min-cut inside $G_i+A+B_q$. We note that since the capacity of $\delta^-(T)$ is exactly $k$, $G_i+A+B_q$ will be $s$-rooted $k$-connected, thus, the edge $(s,q)$ will never be added to $B_q$ and $\delta^-(T)$ remains a $s$-rooted min-cut in the final $G_i+A+B_q$.
\end{proof}

\begin{remark}
    We also note that, the final $G_i+A+B_q$ is $k$-rooted connected. This is because if not, we have added $(s,q)$ to $B_q$ and then $B_q$ will contain an out tree with capacity $+ \infty$ from $s$. In this case, there is no cut with finite size in $G_i+A+B_q$.
\end{remark}

In a nutshell, we state the following lemma as an extension of \Cref{thm:weak-augment}.

\begin{lemma}
    \label{lem:full-augment}
    \Cref{alg:full-augment} guarantees that, it outputs a subpartition $\mathcal{R} = \{R_i\}_{i \in [\ell]}$ of $V$ where there is no edge between each $R_i,R_j \in \mathcal{R}$, a sink set $\mathcal{T} = \{t_i\}_{i \in [\ell]}$, and edge sets $\mathcal{A} = \{A_i\}_{i \in [\ell]}$ where $A_i \subseteq (R_i \cup \{s\}) \times R_i$. Let $G_i$ be the graph obtained by contracting $V-R_i$ to the source $s$. 
    
    Then each $G_i+A_i$ will be $s$-rooted $k$-connected, and any $s$-rooted minimal cut $Z \subseteq E$ with size \textbf{exactly} $k$ becomes an $s$-rooted minimum cut in $G_i+A_i$ where $t_i$ is on the sink side for some $i \in [\ell]$, with probability at least $2^{-\poly(k)}n^{-o(1)}$. 
\end{lemma}

\begin{proof}
    The fact that there is no edge between each $R_i,R_j \in \mathcal{R}$ and each $G_i+A_i$ are $s$-rooted $k$-connected come directly from above discussions. Now we deal with the probability that $Z$ becomes an $s$-rooted minimum cut in $G_i+A_i$. Combining \Cref{lem:full-anal1} and \Cref{lem:full-anal2}, we know this probability is at least \[(k^24^k\log^2 n)^{-O(L)} \times L^{-1} \times 2^{-O(k^4\log k)} \times \Omega((k\theta\log n)^{-1}) = 2^{-O(\sqrt{k\log n} + \sqrt{\log n/k}\log \log n + k^4\log k)},\]
    when we choose $L = \sqrt{\log n/k}$ and $\theta = n^{1/L} = 2^{\frac{\log n}{\sqrt{\log n/k}}} = 2^{\sqrt{k\log n}}$.
\end{proof}

Given the lemma above, we aim to assemble a final edge set $B$ given $\mathcal{R}$ and $\mathcal{A}$.

\begin{itemize}
    \item We let $B_s = \{(s,v)|v \notin \cup_{R \in \mathcal{R}} R\}$ by adding edges from $s$ to every vertex outside any $R\in \mathcal{R}$.
    \item For any edge set $A$ corresponding to $R$, we add $A$ in the original graph $G$ (when there's any edge added from the contracted node to $s$ in $A$, we just add from $s$ itself).
\end{itemize}

Below we want to claim that any rooted min-cut in $G+B$ is corresponding to a rooted min-cut in some $G_i+A_i$. We show this in the following lemma:

\begin{lemma}
    \label{lem:strong-augment}
    Consider the edge set $B = (\cup_{A\in \mathcal{A}} A) \cup B_s$ generated randomly from \Cref{alg:full-augment}, in the graph $G + B$, for each $s$-rooted minimal cut $Z$ in $G$ with capacity \textbf{exactly} $k$, $Z$ becomes an $s$-rooted min-cut with probability $2^{-\poly(k)} \cdot n^{-o(1)}$.
\end{lemma}

\begin{proof}
    We observe that, when any set $T \subseteq R$ for some $\mathcal{R}$, $\rho_{G_i+A_i}(T)$ and $\rho_{G+B}(T)$ will be identical since $G_i+A_i$ can be obtained from $G+B$ by contracting $V-R$ to $s$. Now suppose $Z$ already becomes a $s$-rooted min-cut in $G_i+A_i$ for some $i$  with probability $2^{-\poly(k)} \cdot n^{-o(1)}$ from \Cref{lem:full-augment}. We claim that $Z$ must be an $s$-rooted min-cut in $G+B$.
    
    Suppose towards contradiction that some other $(S,T)$ is a min-cut in $G+B$ with $\rho_{G+B}(T) < k$. First, when $T$ contains any vertex $v$ in $V - \cup R$, since $(s,v) \in B_s$, we know the capacity of the cut is infinity, so this is impossible. Next, when $T$ only contain vertices in one or more component $R \in \mathcal{R}$, we know $\rho_{G+B}(T) = \sum_{R \cap T \neq \emptyset}\rho_{G+B}(R \cap T)$, since there are no edges between any two $R$ components\footnote{Note that, this is the reason that we use $\mathcal{D}_{ij}$ families instead of $\mathcal{C}_i$ families when we combine into $G$.}. Also, inside each component $R$, we have $\rho_{G+B}(R \cap T) = \rho_{G_i+A_i}(R \cap T) \geq k$ since we enforce that $G_i+A_i$ is $k$-rooted connected. Therefore, in this case we also have $\rho_{G+B}(T) \geq k$, which is impossible. Thus, $Z$ is an $s$-rooted min-cut in $G+B$ and the lemma holds.
\end{proof}

Finally, we prove \Cref{thm:strong-augment} as a direct consequence from the above lemma. We note that, the only difference between \Cref{thm:strong-augment} and \Cref{lem:strong-augment} is the condition that $|Z|$ is at most $k$ or exactly $k$. However, this is very easy to resolve.

\begin{proof}[Proof of \Cref{thm:strong-augment}]
    We consider first sample a $t \in [1,k]$, and then construct the edge set $B$ as in \Cref{lem:strong-augment} randomly with parameter $t$ instead of $k$. Now with probability $1/k$ we have $t = |Z|$, and with probability $2^{-\poly(k)} \cdot n^{-o(1)}$, $Z$ becomes an $s$-rooted min-cut in $G+B$ by \Cref{lem:strong-augment}. Therefore, the probability is $(1/k) \cdot 2^{-\poly(k)} \cdot n^{-o(1)}$ is still $2^{-\poly(k)} \cdot n^{-o(1)}$.
\end{proof}


\section{Strong Connectivity Oracle}
\label{sec:oracle}

In this section, we will build an oracle as in \Cref{thm:sccoracle} using \Cref{thm:weak-augment}. For convenience, we state them below before describing our oracle.

\sccoracle*
\weakaugment*

Following \cite{ahi2026maximum}, we now state a stronger version of our oracle, which indicates that we can not only query if the graph is strongly connected, but also $\lambda$-connected as well.

\begin{theorem}
    \label{thm:connoracle}
    Given a directed graph $G$ and a positive integer $k$, there is an oracle that uses $2^{\poly(\lambda+k)}n^{1 + o(1)}$ space, and given any set $F$ of at most $k$ edges, answers if $G \setminus F$ is $\lambda$-connected with high probability in time $2^{\poly(\lambda + k)}n^{o(1)}$. 
\end{theorem}

For convenience, we state the theorem in the above form. In fact, if the actual graph is with connectivity $\mu$, then with trivial modifications the oracle can return $\min(\mu,\lambda)$. 

Still, we start with fixing an arbitrary source $s$. Since $G-F$ is $\lambda$-strongly connected if and only if both $G-F$ and the reverse graph $(G-F)^R$ are $s$-rooted $\lambda$-connected, below, we only consider the problem to check if $G-F$ is $s$-rooted $\lambda$-connected (and we can do this in the reverse graph as well). Our oracle construction will be simply as follows. We first compute $2^{\poly(\lambda+k)}n^{o(1)}$ independent rooted flow-augmentation instances. Then in each instance, suppose we have a graph family $\{G_i\}$ as in \Cref{thm:weak-augment}, we consider the following oracle for $(s,t_i)$ inside $G_i$, for all $i \in [p]$:

\begin{lemma}[\cite{ahi2026maximum}, Theorem 7.3]
    \label{lem:storacle}
    For any directed graph $G$ and a pair of vertices $(s,t)$, let $\mu$ be the $(s,t)$ min-cut size. There exists an $O(\mu n)$-sized oracle that, for any set $F$ of $k \leq \mu$ edges, determines in $O(k^2)$ time whether the $(s, t)$ min-cut size decreases by $k$ upon the failure of $F$.
\end{lemma}

We let $\mu_i$ denote the $(s,t_i)$-connectivity (as well as $s$-rooted connectivity) in each $G_i+A_i$, when $\mu_i \leq \lambda+k$, we construct the oracle $(s,t_i)$ inside $G_i+A_i$, which needs $O((\lambda+k)|V(G_i)|)$ space. Otherwise, we just memorize that $\mu > \lambda+k$ (with $O(1)$ space). Since the total number of vertices in all $G_i+A_i$ is at most $O(n)$, in each instance, we use $O((\lambda+k)n)$ space and in all instances, we still use $2^{\poly(\lambda+k)}n^{1 + o(1)}$ space.

Next, we consider the query phase in each instance when we are given a failed edge set $F$ with $|F| \leq k$. We first show that, there exists a subset $F' \subseteq F$ that can be extended into a $s$-rooted minimal cut in $G$ in the following sense:

\begin{lemma}
    Suppose $Z'$ is an $s$-rooted min-cut in $G-F$, then there exists a subset $F' \subseteq F$, where $Z = F' \cup Z'$ is an $s$-rooted minimal cut in $G$.
\end{lemma}

\begin{proof}
    Consider the $s$-rooted cut $F \cup Z'$, by repeatedly finding edge subsets in $F \cup Z'$ that is still $s$-rooted cuts, we will find an $s$-rooted minimal cut $Z$. We claim that $Z' \subseteq Z$, since we know even $F \cup Z'-z$ is not an $s$-rooted min-cut for every $z \in Z'$. Therefore, we let $F'=Z-Z'$ and finish the proof.
\end{proof}

Now we consider the minimal cut $Z$. When the $s$-rooted connectivity is less than $\lambda$, we know $|Z'|<\lambda$ and thus $|Z| < \lambda + k$. Now since $Z$ is an $s$-rooted minimal cut in $G$, by \Cref{thm:weak-augment}, we know $Z$ becomes a rooted min-cut in some graph $G_i+A_i$ with probability $2^{-\poly(\lambda+k)}n^{-o(1)}$. In that case, suppose we delete $F' \subseteq Z$ from $G_i+A_i$, then the $(s,t_i)$ connectivity will be decreased by exactly $|F'|$. Based on these observations, we complete the query phase.

The query given $F$ is executed as follows: First, we enumerate all $O(2^k)$ different edge subsets $F' \subseteq F$. Then, for each $F'$ and each of the $2^{\poly(\lambda+k)}n^{o(1)}$ instances, there exists at most one set $R_i$ in the subpartition $\mathcal{R}$ that all edges in $F'$ point to. If there exists such a set $R_i$, we consider the oracle built for the corresponding $G_i+A_i$, as in \Cref{lem:storacle}.

Suppose in $G_i+A_i$ we have $\mu_i < |F'| + \lambda$, and the $(s,t_i)$ connectivity is decreased exactly from $\mu_i$ to $\mu_i-|F'|$, then we report the $s$-rooted connectivity is less than $\lambda$. If there is no report throughout the above procedure, we then report the $s$-rooted connectivity is at least $\lambda$. Clearly, the query time is $2^k \cdot 2^{\poly(\lambda+k)}n^{o(1)} \cdot k^2 = 2^{\poly(\lambda+k)}n^{o(1)}$. Finally, we prove the correctness of our query and prove \Cref{thm:connoracle}.

\begin{proof}[Proof of \Cref{thm:connoracle}]
    When the graph $G-F$ is $\lambda$-connected, it is clear that for any graph $G_i+A_i$, after deleting every $F' \subseteq F$, the connectivity will never be dropped to less than $\lambda$. Therefore, the oracle will always return correctly.

    Now suppose $G-F$ is not $\lambda$-connected, i.e., the $s$-rooted min-cut $Z'$ is with size $|Z'| < \lambda$, by the above discussion, in each instance, $Z = Z'+F'$ becomes a rooted min-cut in some graph $G_i+A_i$ with probability $2^{-\poly(\lambda+k)}n^{-o(1)}$. Therefore, when we enumerate the correct $F'$, with probability $2^{-\poly(\lambda+k)}n^{-o(1)}$, in $G_i+A_i$ the connectivity will drop from $\mu_i=|Z|$ to $\mu_i-|F'|=|Z'|< \lambda$ with probability $p = 2^{-\poly(\lambda+k)}n^{-o(1)}$, in which the oracle will return correctly that $|Z'|<\lambda$. Finally, when we have $\log n/p$ (which is still in $2^{\poly(\lambda+k)}n^{o(1)}$) independent flow-augmentation instances, by a standard Chernoff bound as in \Cref{lem:chernoff}, we obtain that the oracle will return correctly with high probability.
\end{proof}

\paragraph{Open Questions.} Currently, our oracle still has an $n^{o(1)}$ factor in both size and query time, which may require further observations to improve. Also, our oracle and rooted flow-augmentation use randomized regimes by sampling. Showing a deterministic variant similar to \cite{kim2022directed} would be interesting. 

\paragraph{AI Disclosure.} We used ChatGPT to assist with the proof of \Cref{lem:DAGdecomp}. The authors verified the correctness of all content in the proof.

\bibliographystyle{alpha}
\bibliography{ref}

\end{document}